\documentclass[12pt,amsmath,amssymb,prd,onecolumn]{revtex4-2}

\usepackage{color}
\usepackage{graphicx}
\usepackage{slashed}
\usepackage{mathtools}

\newcommand{\be}{\begin{equation}}
\newcommand{\ee}{\end{equation}}

\newcommand{\eq}[1]{Eq.~\eqref{eq:#1}}

\renewcommand{\sec}[1]{Sec.~\ref{sec:#1}}
\newcommand{\ssec}[1]{Sec.~\ref{ssec:#1}}

\newcommand{\abs}[1]{\left| #1 \right|}

\newcommand{\pTkT}{p_\perp\!\cdot k_\perp}

\begin{document}

\title{Electron response to radiation under linear acceleration: classical, QED and accelerated frame predictions}
\author{B. M. Hegelich}
\affiliation{Department of Physics, The University of Texas, Austin, 78712, USA}
\author{L. Labun}
\affiliation{Department of Physics, The University of Texas, Austin, 78712, USA}
\author{O. Z Labun}
\affiliation{Department of Physics, The University of Texas, Austin, 78712, USA}
\author{G. Torrieri}
\affiliation{Department of Physics Gleb Watagin, University of Campinas, Campinas, Brazil}
\author{H. Truran}
\affiliation{Department of Physics Gleb Watagin, University of Campinas, Campinas, Brazil}

\date{22 Jan 2022}

\begin{abstract}
A model detector undergoing constant, infinite-duration acceleration converges to an equilibrium state described by the Hawking-Unruh temperature $T_a=(a/2\pi)(\hbar/c)$.  To relate this prediction to experimental observables, a point-like charged particle, such as an electron, is considered in place of the model detector.  Instead of the detector's internal degree of freedom, the electron's low-momentum fluctuations in the plane transverse to the acceleration provide a degree of freedom and observables which are compatible with the symmetry and thermalize by interaction with the radiation field.  General arguments in the accelerated frame suggest thermalization and a fluctuation-dissipation relation but leave underdetermined the magnitude of either the fluctuation or the dissipation.  Lab frame analysis reproduces the radiation losses, described by the classical Lorentz-Abraham-Dirac equation, and reveals a classical stochastic force.  We derive the fluctuation-dissipation relation between the radiation losses and stochastic force as well as equipartitation $\langle p_\perp^2\rangle = 2mT_a$ from classical electrodynamics alone.  The derivation uses only straightforward statistical definitions to obtain the dissipation and fluctuation dynamics.  Since high accelerations are necessary for these dynamics to become important, we compare classical results for the relaxation and diffusion times to strong-field quantum electrodynamics results.  We find that experimental realization will require development of more precise observables. Even wakefield accelerators, which offer the largest linear accelerations available in the lab, will require improvement over current technology as well as high statistics to distinguish an effect.
\end{abstract}

\maketitle

\section{Introduction}
The study of detectors in accelerated states was inspired by the quest to understand Hawking's prediction of thermal radiation from a black hole \cite{Hawking:1974rv}.  A detector undergoing constant acceleration exhibits a thermal excitation spectrum at temperature \cite{Davies:1974th,Unruh:1976db}
\begin{align}\label{eq:TUdefn}
T_a=\frac{a}{2\pi}\frac{\hbar}{c}=\frac{a}{\rm m/s^2} 3.5\times 10^{-25}~\mathrm{eV}.
\end{align}
In each case, the detector is coupled to a massless field which is quantized in the classical spacetime.  The thermal spectrum is manifestly associated with the wavefunctions of the quantized field and can be factored out from the transition probability of the detector.  For this reason, it is frequently said that the massless field viewed by the accelerated detector is in a thermal state \cite{Troost:1977dw,Troost:1978yk,Crispino:2007eb}, as appears to be the case for the massless field in a black hole spacetime \cite{Gibbons:1976es,Candelas:1977zz}. The apparent finding of thermalized behavior in hadronic collisions, including very small systems, has added a phenomenological dimension to these speculations, as the Unruh effect has been advocated as a mechanism under which a coherent classical field configuration dissipates into a thermal distribution in a time-scale parametrically shorter than perturbative expectations \cite{Barshay:1977hc,Kharzeev:2005iz,Castorina:2007eb,Biro:2011ne}.

To understand the apparent thermal state better, we consider a concrete realization: a specific accelerated detector and a consequence of the detector's thermalization that is measurable in the laboratory inertial frame.  Most proposals for experiments involve accelerating electrons \cite{Higuchi:1992we,Higuchi:1992both,Chen:1998kp,Schutzhold:2006gj,Schutzhold:2008zza,Thirolf:2010ozy}, which as the lightest charged particles achieve the highest accelerations.  The problem then is to derive the electron's response to the predicted thermal excitation as well as a dynamical observable measurable in the lab frame.  

As a charged particle undergoing high acceleration, the electron radiates electromagnetically.   The massless photon field should exhibit a thermal distribution in the rest frame of an electron in constant acceleration.  Therefore the electron might reveal an imprint of this thermal bath in some characteristic of its radiation distribution.  This is the basic idea behind two proposals for experiments, based either on the stochastic recoil of the probe particle from the radiation in the accelerated frame \cite{Chen:1998kp} or on correlations in 2-photon emission processes \cite{Schutzhold:2008zza}.

Nonequilibrium quantum theory methods were developed to analyze the real-time dynamics of a classical detector or particle coupled to a quantum field \cite{Anglin:1992ur,Johnson:2000qd,Johnson:2005pf,Galley:2005tj,Galley:2006gs} and from the dynamics compute the radiation \cite{Lin:2005uk,Lin:2006jw,Lin:2006vz} laying to rest questions raised about whether any radiation survives in the far-field.  These real-time calculations are also extended to nonconstant accelerations to test approximations and assumptions of the previous proposals \cite{Doukas:2013noa}.  Perhaps most interesting for experimental observation, the electron transverse momentum ``thermalizes'', i.e. after a sufficiently long time satisfies equipartition at the temperature \eq{TUdefn} \cite{Iso:2010yq},
\begin{align}\label{eq:pTequipartition}
\frac{1}{2m}\langle p_\perp^i p_\perp^j\rangle = \frac{1}{2}T_a\delta^{ij}+\mathcal{O}\left(\frac{a}{m}\right)^2
\end{align}
This equipartition relation is an element of a fluctuation-dissipation relation, apparently consistent with the hypothesis of coupling the transverse momentum fluctuations to a thermal bath at temperature $T_a$, as we discuss below.

Ultimately, progress on understanding the physics content of \eq{TUdefn} must be compared to experiment.  We show that equipartition for transverse dynamics arises in a consistent expansion for small accelerations ($a\hbar/c\ll mc^2$) and small transverse fluctuations ($|\vec p_\perp|\ll mc$) around the approximately constant longitudinal acceleration.  The last condition is the experimental challenge: the acceleration should be approximately constant long enough for transient effects and initial state information to be erased, that is several times longer than the dissipation time.  We obtain the dissipation time from classical radiation theory, finding agreement with previous calculations.  Classical radiation theory also yields the correct noise, proving the fluctuation-dissipation and equipartition theorems.  Since the classical radiation calculation involves only a single scale $a$, $\langle p_\perp^2\rangle$ proportional to $T_a$ is inevitable.  What is nontrivial is the correct numerical factor for the equipartiation relation.  The $\hbar$ in \eq{pTequipartition} arises from the conversion of the classical wave number $k$ of the radiation to the momentum it imparts to the electron/detector upon emission. On the other hand as acceleration approaches $a\hbar/mc^3\to 1$, quantum electrodynamics, can be applied to determine the radiation emitted by the electron.  We evaluate the dissipation time, noise and mean-square transverse momentum using strong-field quantum electrodynamics to quantify the high-acceleration departure from classical predictions of radiation response and the thermal fluctuation-dissipation and equipartition relations.  Before closing, we discuss the timescales in the context of linear accelerator technology and find that both conventional radio-frequency accelerators and wakefield accelerators currently provide gradients that are too small and over too short times to access directly the ``thermalized'' state of an accelerating particle.

\section{Accelerated frame analysis}\label{sec:rindler}

Supposing horizons imply a thermodynamic description of the vacua of a massless field \cite{Bekenstein:1974ax,Bekenstein:1975tw,Israel:1976ur,Sciama:1981hr}, we examine the implications for the dynamics of a probe coupled to such a massless field.  More specifically, lab frame analysis of the two-point correlation function of the radiation field proves it equivalent to a thermal two-point correlator \cite{Troost:1977dw,Anglin:1992ur}.  Concretely of course, we are thinking of describing the dynamics of the electron coupled to the massless photon field in the accelerated, co-moving frame, but the inferences should be applicable more generally.  We refer to the massless field as the radiation field, as in later sections, it is identical with the radiation component of the electromagnetic field.

The simplest consequence is that the expectation value of the energy of the probe degree of freedom should equilibrate at $T_a$,
\begin{align}\label{eq:ETa}
\langle E\rangle =T_a.
\end{align}
This result is straightforwardly applied to models in which the probe degree of freedom is an ``internal'' state $Q$ to which the radiation field couples, as in the Unruh-DeWitt detector \cite{Unruh:1976db}.  In these models, the probability of excitation to an internal state with energy $E$ is given by the usual Bose or Fermi statistics distribution with temperature $T_a$, which implies \eq{ETa}.  

However, most experimental proposals using electrons and electromagnetic radiation involve phase space dynamics in response to the radiation field (electron spin is a notable exception \cite{Bell:1982qr}).  Involving phase space dynamics poses a potential difficulty in that radiation dynamics generally change the acceleration.  Lab frame analysis (\sec{lab}) shows that radiation causes the acceleration in a general state to decrease to a well-defined non-zero minimum.  This dynamic will shortly be derived in the accelerated state as well.  Clearly we must assume for the moment--and justify \emph{a posteriori}--that the accelerated state can be treated as quasi-stationary, so that the decay is much slower than the dynamics we are considering and the acceleration and temperature can be considered approximately constant.  Without the quasi-stationary approximation, applying a thermodynamic description would be nonsense.  

Additionally, for the interaction of the probe (electron) with an accelerated frame radiation field to be described by classical thermodynamics, the temperature must be much less than the mass of the probe, $T_a\ll m$.  Otherwise, the radiation field would have enough energy to probe the internal structure of the probe and create electron-positron pairs.  This condition is equivalent to the lab frame condition that the probe particle must have negligible recoil from interactions with the radiation field and supports the \emph{a posteriori} justification that the accelerated state is at least quasi-stationary.

To use the accelerated electron as the probe and its radiation as a signal accessible in the lab frame, we need a degree of freedom which interacts with the radiation field in such a way that the dynamics can be computed in both the lab frame and the accelerated frame.  The simplest choice, if it exists, is an observable invariant under the change in frame.  Since any point on the accelerated trajectory is related to the lab frame by a boost (and the accelerated trajectory itself is boost invariant), we are looking for an observable invariant under boosts along the direction of the acceleration.  Such longitudinal boosts leave the transverse directions invariant, so observables describing dynamics in the transverse plane should be equal whether computed in lab or accelerated frame.  Equality of observables has been verified explicitly for the probability of photon emission per unit transverse momentum by Refs. \cite{Higuchi:1992both}.

Therefore, we can investigate $(\vec x_\perp,\vec p_\perp)$ dynamics of the probe to seek effects of the thermal state of the radiation field.  The first inference is that equipartition \eq{ETa} should be applicable to the transverse kinetic energy.  Since we are limited to the locally nonrelativistic regime $T_a\ll m$ (in the instantaneus co-moving frame the motion is non-relativistic for much longer than equilibration time defined below), we have $E_\perp\simeq p_\perp^2/2m$ 
\begin{align}\label{eq:pTsquaredTa}
\frac{1}{2m}\langle p_\perp^i p_\perp^j\rangle = \frac{1}{2}T_a\delta^{ij},
\end{align}
The difference between this statement and \eq{pTequipartition} is that this has been obtained from general reasoning about the accelerated state, whereas \eq{pTequipartition} was obtained from a lab frame calculation \cite{Iso:2010yq}.  The relativistic correction to the kinetic energy would imply a correction to the right hand side of $T^2/4m^2$, which we can compare to $T/m$ corrections from other calculations.

A second inference is to recall that under these conditions the dynamics of a heavy probe coupled a thermal bath are described by Brownian motion.  Specifically, according to \eq{pTsquaredTa} we have a heavy particle with momentum $p_\perp\sim \sqrt{mT_a}$ and velocity $v_\perp\sim \sqrt{T_a/m}\ll 1$.  Since $p_\perp\gg T$ and collisions with momentum transfer $\Delta p_\perp\sim T$ are rare, many collisions are required to significantly change the momentum.  Therefore, we can model the interaction as dominated by dissipation and uncorrelated kicks.  The dynamics are then described by a (macroscopic) Langevin equation, defined for the transverse momentum \cite{reif1965fundamentals},
\begin{align}\label{eq:langevin}
\frac{dp^i}{ds}=-\frac{1}{\tau_D}p^i+\xi^i, \qquad \langle \xi^i(s)\xi^j(s')\rangle=\kappa\delta(s-s')\delta^{ij},
\end{align}
where $\tau_D$ is the dissipation (or relaxation) time and $\xi^i$ is a classical random variable describing the stochastic force.  The time variable $s$ in the accelerated frame is the proper time of the accelerated probe.  The dissipation time $\tau_D$ is the timescale for the exponential decay of correlations, including initial data.  For a thermal bath, the dynamics of $\xi^i$ are completely determined by its 2-point function, which being a $\delta$-function in time represents white noise and has no higher order correlations.  $N_d\kappa$ is the mean-square momentum transfer per unit time.  The number of spatial dimensions $N_d=2$ in our case but we keep it as an explicit factor to highlight how various thermodynamic relations are affected by the conversion from usual 3-dimensional dynamics to 2 dimensions.  

The relationship between momentum loss and diffusion is described by a fluctuation-dissipation theorem, which follows from the general analysis of thermal equilibrium between the probe and the thermal bath \cite{reif1965fundamentals}.  Integrating \eq{langevin} leads to the mean square momentum 
\begin{align}
\langle p^2_\perp\rangle\underset{t\gg\tau_D}{\longrightarrow}\frac{N_d}{2}\tau_D\kappa
\end{align}
Since equilibration in the long time limit $t\gg\tau_D$ requires \eq{pTsquaredTa}, we obtain the fluctuation dissipation relation 
\begin{align}\label{eq:kappatauTreln}
2mT_a=\kappa\tau_D
\end{align}
which is independent of $N_d$.  Since $\tau_D$ is the timescale to erase initial conditions, it is also the minimum (proper) duration of the quasi-constant period of acceleration required for these thermal dynamics to become dominant (see Ref \cite{Doukas:2013noa} for calculations of equilibration times for nonconstant acceleration).

Integrating the momentum to obtain the mean square transverse displacement yields
\begin{align}
\langle \Delta x_\perp(t)^2\rangle=2N_d\frac{T_a}{m}\tau_D t
\end{align}
and comparison to the definition of the diffusion constant
\begin{align}
\langle \Delta x_\perp^i(t)\Delta x_\perp^j(t)\rangle=2Dt\delta^{ij}
\end{align}
shows that
\begin{align}\label{eq:DTmtaureln}
D=\frac{\kappa}{2m^2}\tau_D^2=\frac{T_a}{m}\tau_D.
\end{align}
The latter equality has the form of an Einstein relation $D\propto T$, modulo temperature dependence of $\tau_D$, which we will find is essential.

Thus we have 3 characteristic quantities for the fluctuation and dissipation dynamics, and 2 relations determined by thermodynamics.  We need to compute at least one of these from the microscopic theory describing collisions between the probe and the thermalized particles.  Naively, it appears we could compute the mean square momentum transfer per unit time from a standard finite temperature field theory in the limit of a heavy scatterer (e.g. as in Ref. \cite{Moore:2004tg}), but as we discuss below, such calculations will appear in disagreement with the present results since they results in $\kappa\propto e^4$.

\section{Lab frame analysis}\label{sec:lab}
From the lab frame, the electron is undergoing constant acceleration.  Fluctuations in the transverse momentum converge to a steady state in which the mean square momentum is proportional the temperature $T_a$, as would be expected for thermalization \cite{Iso:2010yq}.  Verifying this steady state would provide evidence for the thermal character of the interaction of the electron with the radiation field.  In this section, we show this apparently thermal character is derived from classical electromagnetic theory.  We compare the classical approach to the quantum dynamical formalisms of Refs. \cite{Galley:2005tj,Galley:2006gs,Iso:2010yq}.  As the effect of the accelerated state thermalization is expected to become more important for high accelerations, we compute the same observables in quantum electrodynamics in order to obtain corrections proportional to $T/m\sim a/m$.

\subsection{Classical electrodynamics}\label{ssec:cled}
Classical electrodynamics predicts that any accelerating charged particle radiates, in general causing the particle to lose energy.  We recall some of the basic equations here for comparison to the approaches below.  The starting point, the classical action, is
\begin{align}\label{eq:Scl}
S&=-m\int \sqrt{u^\mu u_\mu} d\tau-\int \frac{1}{4}F^{\mu\nu}F_{\mu\nu} d^4x-\int j^\mu(x) A_{\mu}(x) d^4x
\end{align}
where the classical point-particle current is
\begin{align}\label{eq:jcldefn}
j^\mu=-eu^\mu\delta^4\left(x-\xi(\tau)\right)
\end{align}
with $u^\mu=p^\mu/m$ the electron 4-velocity and $\xi(\tau)$ its trajectory.  Constant, linear acceleration is provided by a homogeneous and static electric field, and as usual we are implicitly splitting the electromagnetic field into a classical, external field, $A_{\mu}^{\rm cl}$ which is not perturbed by the probe particle, and a dynamic radiation field $A_{\mu}^{\rm rad}$ which is sourced by the particle dynamics.  Integrating the Lorentz force equation for a general electron momentum, we find the 4-velocity $u^\mu$ and trajectory $\xi^\mu$ recalled in Appendix \ref{app:dNclcalc}, and the magnitude acceleration in a constant electric field is
\begin{align}
a^\mu a_\mu=-\frac{|eE|^2}{m^2}\frac{p_\perp^2+m^2}{m^2},
\end{align}
which is equal to $|eE|/m$ only for $p_\perp=0$.  The minus sign is due to the 4-acceleration being spacelike.  Any nonvanishing transverse momentum perturbs the acceleration from the naive value.  However, even as $p_\perp^2$ acquires a nonvanishing expectation value due to radiation, its magnitude is consistent with the implicit expansion in $a/m\sim T_a/m$.

Computing the momentum flux of the $A_\mu^{\rm rad}$ field through a sphere at infinity provides the rate of 4-momentum radiated by the electron \cite{coleman1982classical,nikishov1985problems},
\begin{align}\label{eq:radiated4momentum}
dP^\mu_{\rm rad}=-\frac{1}{2}\mathrm{sgn}(k_0)\delta(k^2)k^\mu j(k)\cdot j(k)^* \frac{d^4k}{(2\pi)^3},
\end{align}
where $k^\mu=(\omega_k,\vec k)$ with $|\vec k|=2\pi/\lambda$ is the wave 4-vector of the radiation field. After inserting the classical trajectory and integrating, one finds the usual Larmor formula,
\begin{align}\label{eq:Larmor}
\frac{dP^\mu_{\rm rad}}{d\tau}=-\frac{e^2}{6\pi}a^\nu a_\nu u^\mu=-\frac{dp^\mu_{\rm loss}}{d\tau}
\end{align}
which is manifestly positive.  The trajectory and other supporting calculations are found in Appendix.  In this construction, this momentum loss is not incorporated in the solution of the trajectory entering the current.  It is added to the Lorentz force equation to obtain a radiation-corrected equation of motion, known as the Lorentz-Abraham-Dirac (LAD) equation,
\begin{align}\label{eq:LAD}
\frac{dp^\mu}{d\tau}&=F^\mu_{\rm ext}+\frac{e^2}{6\pi m}\left(p^\mu \left(\frac{du^\mu}{d\tau}\right)^2+\frac{da^\mu}{d\tau}\right),
\end{align}
where $F^{\mu}_{\rm ext}$ is the driving force, here the Lorentz force $F^\mu_{\rm ext}\to qF^{\mu\nu}u_\nu$. 
 The damping timescale due to radiation emission is derived from 
\begin{align}\label{eq:dampingtime}
\frac{1}{\tau_D}\simeq& \frac{1}{E}\frac{dP^0_{\rm rad}}{d\tau} =\frac{e^2}{6\pi m}a^2=\tau_e\frac{a^2}{c^2},
\end{align} 
restoring powers of $c$ in the last equality.  $\tau_e$ is the timescale arising with the LAD,
\begin{align}\label{eq:taue}
\tau_e=\frac{e^2}{6\pi \epsilon_0 mc^3}\simeq 6.24\times 10^{-24}\,\mathrm{s},
\end{align}
related in turn to the Larmor radiation rate, but is not the timescale associated with the dissipation of the charged particle's energy.  As expected, the dissipation time $\tau_D$ is inversely proportional to the acceleration and is classical.

Considering the acceleration exactly constant $da^{\mu}/d\tau=0$ and ignoring the second term in parentheses in \eq{LAD} leaves an equation of the Langevin form $\frac{dp^\mu}{d\tau}=F^{\mu}_{\rm ext}-p^\mu/\tau_D$.
However, the second term is required in the equation of motion to conserve the norm of the 4-momentum $p^2=m^2$, and therefore arises in any consistent derivation of dynamics from the classical electrodynamic action.  Consequently the complete two-term LAD correction is obtained from a more rigorous linearization of the response of the particle to its radiation field \cite{Johnson:2000qd,Galley:2005tj,Galley:2006gs} together with \eq{dampingtime} \cite{Iso:2010yq}.  

Now by interpreting the classical results in terms of photon emission, we can compute higher order moments of the radiation, such as the mean square momentum transfer, for comparison to the accelerated frame.  To start, the number of photons emitted is determined(estimated) from the radiated 4-momentum as 
\begin{align}\label{eq:dNcl}
dN_\gamma^{\rm cl}&=\frac{dP^0_{\rm rad}}{k^0} =-\frac{1}{2}\mathrm{sgn}(k_0)\delta(k^2) j(k)\cdot j(k)^* \frac{d^4k}{(2\pi)^3}.
\end{align}
To determine how fluctuations in the radiation contribute to the electron dynamics, we need the mean square transverse momentum transfer from photon emission
\begin{align}\label{eq:kappacldefn}
N_d\kappa_{\rm cl}=\frac{d}{d\tau}\langle \Delta p_\perp^2\rangle=\int d^2k_\perp \frac{dN_\gamma^{\rm cl}}{d\tau d^2k_\perp}\Delta p_\perp^2.
\end{align}
where $\Delta p_\perp$ is the momentum transfer during the radiation process.  Clearly the $\delta$ function in $dN_\gamma^{\rm cl}$ \eq{dNcl} reduces one of the $k$ integrals but to obtain a rate per unit (proper) time $d\tau$, we must convert from the longitudinal momentum $dk_z$.  

There are two ways to obtain the emission rate differential in time and transverse momentum.  The first method is to calculate from first principles.  The Fourier transformed current $j^\mu(k)$ in \eq{dNcl} involves an integral over $t$, but instead of evaluating each Fourier integral individually (as in Refs. \cite{Higuchi:1992both,Biro:2011ne}) the current correlator $j(k)\cdot j(k)^*$ can be written in terms of average and relative electron rapidity $y$, related to proper time by $y=a\tau/c$.  Due to the boost invariance of the source, emitted photon rapidity is determined only by the average rapidity.  Integrating over photon rapidity therefore eliminates dependence on average rapidity, yielding the emission rate per unit transverse momentum per unit rapidity of the source.  This procedure is described in detail in Appendix \ref{app:dNclcalc}.

The second method is perhaps more transparent and utilizes the same symmetry of the problem, but relies on a semiclassical estimate of the region of the $t$ integration contributing for each photon wavenumber $k$.  Due to the boost invariance of the source, the fully differential emission probability 
\begin{align}
dN_{\rm cl}=\frac{dP^0_{\rm rad}}{k_0}
=\frac{e^2m^2e^{\pi\kappa_\perp}}{4\pi^3(eE)^2}\left(\left(\frac{E^2_\perp}{m^2}\left(1-\frac{\kappa_\perp^2}{\kappa_\Vert^2}\right)-1\right)K_{i\kappa_\perp}(\kappa_\Vert)^2+\frac{E_\perp^2}{m^2}\left(K'_{i\kappa_\perp}(\kappa_\Vert)\right)^2\right)\frac{d^3k}{2k_0},
\end{align}
depends on the photon longitudinal wavenumber $k_z$ only in the phase space factor $dk_z/2k_0$.  Consequently, the $k_z$ integral diverges logarithmically, as evidenced by the result for a finite interval,
\begin{align}
\int_{-k_z^{\rm max}}^{k_z^{\rm max}}\frac{dk_3}{2k_0}=\mathrm{asinh}\frac{k_z^{\rm max}}{k_\perp}.
\end{align}
Now saddlepoint analysis of the Fourier integral of the current correlator $j(k)\cdot j(k)^*$ corroborates the reasoning in the previous paragraph: the dominant contribution to probability comes from a region of the source's trajectory determined by its average momentum, $\tau_{\rm s.p.}=(p_z+p_z')/2eE$, with width $\delta \tau_{\rm s.p.}=|p_z-p_z'|/eE=|k_z|/eE$.  It follows that the integrations over $\tau$ and $k_z$ are equivalent as they are for spontaneous pair creation \cite{Cohen:2008wz}, with the interval of $k_z$ covered corresponding (up to scaling) to the interval of $\tau$ covered,
\begin{align}
\mathrm{asinh}\frac{k_z^{\rm max}}{k_\perp}\simeq \ln\frac{2k_z^{\rm max}}{k_\perp}=\ln\frac{eEt}{m}+\mathrm{const.}
\end{align}
As the dependence is logarithmic, the differential relation is known only up to a constant scaling, 
\begin{align}\label{eq:dkzdtaureln}
\frac{dk_z}{k_0}=C\frac{eE}{mc^2}d\tau.
\end{align}
No $\hbar$ appears since $eE/m$ has units of acceleration.  Comparison with the first-principles calculation (Appendix \ref{app:dNclcalc}) checks that the constant scaling factor is $C=1$.

Applying the variable change \eq{dkzdtaureln}, we obtain in the limit of zero electron transverse momentum
\begin{align}\label{eq:classicaldNd2kT}
\frac{dN^{\rm (cl)}_\gamma}{d^2k_\perp d\tau}=\frac{e^2}{4\pi^3\epsilon_0}\frac{1}{a}\left(K_1(k_\perp/a)\right)^2,
\end{align}
where $K_\nu(z)$ is a modified Bessel function of the second kind.
No $\hbar$ appears in the classical emission probability $N_\gamma^{\rm cl}/d\tau d^2k_\perp$.  By itself, the second moment of the transverse wave number, $\langle k_\perp^2\rangle=\int k_\perp^2(dN/dtd^2k_\perp)d^2k_\perp$, also remains a classical quantity.  However to obtain the mean square momentum transfer to the electron per unit time, we must multiply the wave vector $k$ by $\hbar$ to obtain the correct units, $\Delta p_\perp=\hbar k_\perp$.  In fact, we need only one power of $\hbar$ since $kdN\propto dE$ \eq{dNcl}.  
The modified Bessel function diverges like $K_\nu(z)\sim z^{-\nu}$ for small $z$, so the transverse wavenumber approaches the conformal limit at small $k_\perp$, like that of a free unaccelerated charge.  The distribution \eq{classicaldNd2kT} is exponentially suppressed at high $k_\perp$, with a temperature-like parameter proportional but not equal to $T_a$ \cite{Biro:2011ne} (because $a$ is the only scale in the classical radiation problem).  The integal of the modified Bessel functions is analytic and yields a constant with the result
\begin{align}\label{eq:kappaclresult}
\kappa_{\rm cl}&=\frac{d\langle \Delta p_\perp^2\rangle}{d\tau}=\frac{1}{\hbar}\int d^2k_\perp (\hbar k_\perp)^2 \frac{dN^{\rm cl}_\gamma}{d\tau d^2k_\perp} \notag\\
&=\frac{e^2}{6\pi^2\epsilon_0}\frac{\hbar}{c^6}a^3.
\end{align}

These properties of the emission probability support a picture of the radiation dynamics like that in the accelerated frame, even without the hypothesis of a thermal bath.  Specifically, since collisions with small momentum transfer are frequent, causing dissipation known as radiation reaction \eq{LAD}, and collisions with momentum transfer $\Delta p_\perp\sim T$ are rare, many collisions are required to significantly change the momentum and we might model the interaction as dominated by dissipation and uncorrelated kicks.  We could therefore hypothesize a generalized Langevin equation for the transverse dynamics, with the LAD radiation loss term replacing the dissipation term $-p^i/\tau_D$ in \eq{langevin},
\begin{align}\label{eq:classicalLangevin}
\frac{dp^i}{d\tau}=F_{\rm ext}^i+\frac{e^2}{6\pi m}\left(p^i \left(\frac{du^i}{d\tau}\right)^2+\frac{da^i}{d\tau}\right)+\xi^i,\quad \langle \xi^i(\tau)\xi^j(\tau')\rangle=\kappa_{\rm cl}\delta(\tau-\tau')\delta^{ij}.
\end{align}
The stochastic force $\xi^i$ has the same form as for the Langevin equation because the kicks are assumed to be uncorrelated.  In principle, computing higher order correlation functions of the radiation, we should find higher order correlations in the noise, but these are suppressed by the coupling.  Combining \eq{kappaclresult} with \eq{dampingtime}, we find 
\begin{align}\label{eq:kappacltauT}
\kappa_{\rm cl}\tau_D=2m\frac{\hbar a}{2\pi c}=2mT_a,
\end{align}
and integrating \eq{classicalLangevin} would lead to $\langle p_\perp^2\rangle=2mT_a$ upon using \eq{kappacltauT}.  According to \eq{DTmtaureln} the diffusion constant would be
\begin{align}\label{eq:Dcl}
D_{\rm cl}=\frac{\kappa_{\rm cl}\tau_D^2}{2m^2}=\frac{3\epsilon_0}{e^2a}\hbar c^4,
\end{align}
with the $\hbar$ coming from $\kappa$.
While the justification for \eq{classicalLangevin} is a bit hand-waving at this point, we can derive it rigorously with guidance from a different but closely related approach to the electron-radiation interaction, namely considering $A_{\mu}^{\rm rad}$ as a quantized photon field.

\subsection{Quantized photon dynamics}\label{ssec:quantphoton}

The original black hole and accelerated detector problems were formulated as the interaction of a classical object or detector with a quantized field, and therefore it has been natural for most authors to study the dynamics of the quantized radiation field, which is easily compared between frames.  However for the massless and uncharged photon field, it turns out that calculations of the radiation distribution with a quantized radiation field from a classical point source are equivalent to calculations within classical radiation theory \cite{Higuchi:1993fn}.

The equivalence is highlighted by computing the probability of photon emission.  The action is the same as the classical action \eq{Scl}, modulo a gauge fixing term which we do not need for the tree-level calculations here.  Only the photon is quantized.  The probability of photon emission differential in photon wave number is related to the squared matrix element for photon emission,
\begin{align}\label{eq:photonemissionM}
dW&=\sum_{\epsilon,\epsilon'}|\mathcal{M}|^2\frac{d^3k}{(2\pi)^32|\vec k|}\\
\mathcal{M}&=\int d^4x \langle \vec k,\vec\epsilon|j(x)\cdot \hat{A}(x)|0\rangle=\int d^4x (j(x)\cdot\epsilon) e^{-i|\vec k|t+\vec k\cdot \vec x}.
\end{align}
The current is classical, so the matrix element is straightforwardly evaluated in terms of plane waves and the polarization vector $\epsilon^\mu$ of the photon field, which satisfies $k\cdot\epsilon=0$.  Rewriting the photon wavenumber phase space using a $\delta(k^2)$, we have
\begin{align}
dW=\sum_{\epsilon,\epsilon'}\int d^4x\int d^4x'(\epsilon\cdot j(x))(\epsilon'\cdot j(x'))e^{-ik(x-x')}]\frac{1}{2}\mathrm{sgn}(k_0)\delta(k^2)\frac{d^4k}{(2\pi)^3}.
\end{align}
Then using the usual polarization sum identity $\sum_{\epsilon,\epsilon'}\epsilon_\mu\epsilon'_\nu=-g_{\mu\nu}$ and the definition of the Fourier transform, we are back to the classically obtained expression \eq{dNcl}.  

Neither classical radiation theory nor the quantized radiation field have the power to compute all observables.  While \eq{dNcl} or \eq{photonemissionM} can be used to compute the spectrum and moments of the photon distribution, they cannot compute the radiation intensity, which relies on considering the emission as a continuous process and the radiation as a continuous field.  Extensions of the quantized photon approach using nonequilibrium quantum theory methods enable investigation of the system-environment separation and the conditions and dynamics of decoherence.  Such more powerful methods are necessary to determine more quantitatively when the intuitive picture of dynamics obtained here is valid.

Sacrificing some rigor for clarity, we can simplify the calculation of the feedback of the radiation on the classical source to obtain a generalized Langevin equation of the form \eq{classicalLangevin}.  The leading order equation of motion for the current is the Lorentz force,
\begin{align}
\frac{dp^\mu}{d\tau}=qF^{\mu\nu}u_\nu,
\end{align}
which if we separate $F^{\mu\nu}$ into an external field and the photon field, $F^{\mu\nu}=F^{\mu\nu}_{\rm ext}+\hat{F}^{\mu\nu}$, can be rewritten
\begin{align}\label{eq:LFextandphoton}
\frac{dp^\mu}{d\tau}=F^{\mu}_{\rm ext}+q\hat{F}^{\mu\nu}u_\nu, \qquad F^{\mu}_{\rm ext}\equiv qF^{\mu\nu}_{\rm ext}u_\nu.
\end{align}
The external field generates the leading order classical trajectory, around which we will perturb.  From the action, we construct an iterative solution for the photon field $\hat{A}^\mu$.  With the Lorenz gauge condition 
\begin{align}
\partial_\mu \hat{A}^\mu=0,
\end{align}
the equation of motion for $A^\mu$ is the Maxwell equation,
\begin{align}\label{eq:MaxwellAmu}
j^\nu=\partial_\mu F^{\mu\nu}=\partial^2 A^\nu
\end{align}
with $j^\nu$ the classical current \eq{jcldefn}.  

The general solution to \eq{MaxwellAmu} is $A^\mu(x)=A^\mu_{\rm h}(x)+A^\mu_{\rm inh}(x)$, the sum of a homogeneous solution $A^\mu_{\rm h}$, which brings in the vacuum (free-field) dynamics of the photon, and an inhomogeneous solution $A^\mu_{\rm inh}$, which brings in the excitation of the photon field by the classical source current.  Assuming the initial state of the radiation field is gaussian, consistent with a free field state uncoupled to the charge, the homogeneous solution contributes a stochastic field with a nominally classical probability distribution, whereas the inhomogeneous solution contributes the history-dependent dissipation \cite{Johnson:2000qd,Calzetta:2008iqa}.  The reason for this separation is analyticity: the propagator for the radiation field can be separated into real and imaginary parts, which under causal construction devolve respectively to the Hadamard and retarded propagators.

Formally, we obtain the same result by inserting the homogeneous solution and inhomogeneous solution into \eq{LFextandphoton}  \cite{Iso:2010yq}.  The homogeneous solution, solving $\partial^2A=0$, is a complete set of plane waves,
\begin{align}A^{\mu}_{\rm h}(x)=\int \frac{d^3k}{(2\pi)^3}\frac{1}{\sqrt{2k^0}}\left(\epsilon^\mu_k a_ke^{-ik^\nu x_\nu}+\epsilon^{*\mu}_ka^\dag_ke^{ik^\nu x_\nu}\right),\end{align}
satisfying the usual on-shell condition $k^0=|\vec k|$.  The polarization vectors satisfy $k\cdot\epsilon_k=0$ and the mode functions $a_k,a^\dag_k$ are classical amplitudes.  The inhomogeneous solution is constructed from the retarded Green's function,
\begin{align}
A^{\mu}_{\rm inh}(x)=\int d^4x' G_R(x,x')j^\mu(x')
\end{align}
where the Green's function satisfies
\begin{align}\label{eq:GreensfnEoM}
\partial^2_xG_R(x,x')=\delta^4(x-x') 
\end{align}
and.  With this Ansatz for $A^\mu(x)$, using the $\delta$ functions in \eq{jcldefn} to reduce the $x'$ integral and regularizing the singular contributions from the $\tau'\to\tau$ limit \cite{Galley:2005tj,Galley:2006gs}, we obtain 
\begin{align}
\frac{dp^\mu}{d\tau}=F^{\mu}_{\rm ext}+q(\partial^\mu\hat{A}_{\rm h}^\nu-\partial^\nu\hat{A}_{\rm h}^\mu)u_\nu+\frac{e^2}{6\pi m}\left(p^\mu \left(\frac{du^\mu}{d\tau}\right)^2+\frac{da^\mu}{d\tau}\right).
\end{align}
Like the Langevin equation, this equation describes the dynamics of an observable; physical quantities are expectation values of the observable and its moments.  The expectation value defines the contribution of the stochastic field $\hat A_{\rm h}$, which has the properties of a noise field $\langle \hat A_{\rm h}(x)\rangle=0$ and must be symmetrized before evaluating the two-point function $\langle\hat A_{\rm h}(x)\hat A_{\rm h}(y)\rangle\to\frac{1}{2}\langle\{\hat A_{\rm h}(x),\hat A_{\rm h}(y)\}\rangle$ corresponding to the Hadamard propagator arising in the more rigorous derivation.

To investigate small transverse fluctuations, we linearize around the zeroth order solution, $p^\mu=p^\mu_{(0)}+\delta p^\mu$, that satisfies the external force $dp^\mu_{(0)}/d\tau-F^\mu_{\rm ext}=0$.  
In agreement with the classical estimate, the solution to the stochastic equation of motion for transverse motion shows the damping time scale for transverse dynamics to be $\tau_D=c^2/a^2\tau_e$ identical to \eq{dampingtime}.  Further, it is verified by explicit calculation that mean square momentum converges after long times $\tau\gg\tau_D$ to (Eq. 5.15 of Ref. \cite{Iso:2010yq})
\begin{align}
\frac{1}{2m}\langle \delta p^i\delta p^j\rangle=\frac{1}{2}T_U\delta^{ij}\left(1+\mathcal{O}\left(\frac{a^2}{m^2}\right)\right).
\end{align}
By analysis similar to the Langevin dynamics, we obtain the diffusion constant from the long time dynamics of the mean square transverse displacement.  The result is
\begin{align}
D=\frac{3}{e^2a}
\end{align}
 in agreement with \eq{Dcl}\,\footnote{Note however that in \cite{Iso:2010yq}, an incorrect definition of $D$ in Eq. 3.23 leads to a result differing by a factor 2 from ours.}.  The mean square momentum transfer $\kappa$ is not explicitly defined as such in this approach, but it can be read off from from the calculation of the field correlator (Eq. 3.11 of \cite{Iso:2010yq}) and multiplying by factors of $e^2$ (for the coupling) and 2 (for the 2 polarizations of the photon)
\begin{align}
\kappa &=\frac{e^2a^3}{6\pi^2} \label{eq:kappaviso}
\end{align}
in agreement with \eq{kappaclresult}.

Although this approach yields the same observable results as classical radiation theory, it provides a more rigorous basis for introducing the Langevin dynamics and understanding its origin in neglecting higher order correlations in the radiation field.

\subsection{Quantum electrodynamics}\label{ssec:QED}

To obtain corrections at high acceleration $a/m\to 1$ we must start from a theory that accounts for recoil from photon emission.  The electron must be quantized in order to conserve momentum at each emission.  As the constant electric field generates dynamics identical to uniform acceleration, we quantize the electron in the classical gauge potential $A^{\mu}_{\rm cl}=-eEt\delta^\mu_3$ corresponding to a homogeneous and static electric field in the $\hat z$ direction.  The time-dependent gauge is chosen for this time-dependent problem.  The hard work of constructing wavefunctions and simplifying the matrix element has been done \cite{nikishov1985problems} and salient aspects of the calculation reviewed in Appendix \ref{app:sfQED}.  The fully differential probability, at $p_\perp=0$, is
\begin{align}\label{eq:dWd3kMsquared}
dW&=\frac{d^3k}{(2\pi)^32k_0}\frac{1}{2}\sum_{\substack{\sigma,\sigma' \\ \epsilon,\epsilon'}}\int\frac{d^3p'}{(2\pi)^32E_{p'}}\left|\mathcal{M}[e_{\vec p}\to e_{\vec p'}\gamma_{\vec k}]\right|^2 \notag \\
&\equiv \frac{d^3k}{k_0}\frac{1}{|eE|}w(k_\perp^2,|eE|) \\
w(k_\perp^2,|eE|)&=\frac{\alpha}{2\pi} \frac{e^{-\pi\frac{k_\perp^2}{eE}}}{(1-e^{-\pi\frac{k_\perp^2+ m^2}{eE}})2\sinh(\frac{\pi m^2}{eE})}\frac{k_\perp^2}{k_\perp^2+m^2}
\left[\left(2+\frac{k_\perp^2}{m^2}\right)|\Psi'|^2+|\Psi|^2+2\mathrm{Re}[\Psi'\Psi^*]\right] \notag \\
\Psi&=\Psi\!\left(\frac{im^2}{2eE},1-\frac{ik_\perp^2}{2eE};\frac{-ik_\perp^2}{2eE}\right)
\end{align}
where $\Psi(a,b;z)$ is the second confluent hypergeometric (see \eq{hypergeomreln}) and the prime denotes differentiation with respect to the argument $z$, $\Psi'(a,b;z)=d\Psi/dz$.  For notational brevity, we have suppressed the $\hbar$s in this expression. From this, we need to compute two quantities for comparison, the dissipation time $\tau_D$ and the mean-squared momentum transfer per unit time $\kappa$.  

The first, $\tau_D$ encounters the difficulty pointed out in the previous section: in quantized radiation dynamics, we do not have a definition of continuous momentum flux in the radiation field, since it is composed of the probabilities of finding quanta in a given mode.  To obtain a definition of the energy loss rate, we extend the semiclassical analysis of \ssec{cled}.  The discussion above \eq{dkzdtaureln} showed that the probability of emission in a given $k_z$ mode is dominated by a saddle-point on the electron's trajectory determined by the electron's momentum.  Therefore we can say that the energy lost over a given finite interval is given by integrating over the corresponding $k_z$ (and all $k_\perp$) and dividing by the duration of the interval \cite{nikishov1985problems},
\begin{align}
\frac{\Delta E}{\Delta t}&=\int d^2k_\perp \frac{1}{\Delta t_{\rm s.p.}}\int_{-k_z^{\rm max}}^{k_z^{\rm max}}\! dk_z \frac{1}{|eE|}w(k_\perp^2,|eE|) \\ \notag
\Delta t_{\rm s.p.}&=\frac{m}{eE}\frac{2k_z^{\rm max}}{k_\perp}.
\end{align}
Since this is an estimate expected to be valid to within a constant of order unity, we introduce a constant in the time interval $\Delta t\to C_t\Delta t$ with which we match to the classical result.  Since $k^0dN_\gamma$ is independent of $k_z$, the integral yields $2k_z^{\rm max}$, which cancels with the same factor in $\Delta t_{\rm s.p.}$.  The result is 
\begin{align}
\left.\frac{\Delta E}{\Delta t}\right|_{\rm QED}&=
\frac{1}{C_tm}\int d^2k_\perp k_\perp w(k_\perp^2,|eE|).
\end{align}

To determine the constant $C_t$, we take the classical limit $\hbar\to 0$.  The limit is clarified by writing all parameters in terms of the dimensionless quantities $k_\perp \ell_a$ and $\lambda_e/\ell_a$ where $\ell_a=c^2/a=m_ec^2/eE$ is the length scale associated with the classical acceleration and $\lambda_e=\hbar/m_ec$.  Thus the $\hbar\to 0$ limit is manifestly the limit of a point-like electron, i.e. the Compton wavelength vanishes relative to the acceleration length scale, $\lambda_e/\ell_a=(\hbar/m_ec)/(m_ec^2/eE)\to 0$.  As expected from the Euler-Heisenberg effective action, quantum effects become important as $a/m\sim 1$ \cite{Heisenberg:1936nmg}, which is equivalent to the electric field approaching the ``critical field'' $eE\sim m_e^2c^3/\hbar$.  Using Eq 8.14 of  Ref. \cite{nikishov1985problems}, the limit is
\begin{align}
\lim_{\hbar\to 0}\left.\frac{\Delta E}{\Delta t}\right|_{\rm QED}&=\frac{e^2}{2\pi^2}\frac{1}{C_t a}\int_0^\infty \left(K_1(k_\perp/a)\right)^2 k_\perp^2 dk_\perp
=\frac{9\pi}{32C_t}\frac{e^2}{6\pi}a^2,
\end{align}
which fixes $C_t=9\pi/32$.  
The relaxation time is then defined paralleling the classical estimate \eq{dampingtime}, 
\begin{align}\label{eq:QEDtauD}
\tau_{Dq}^{-1}=\frac{1}{E}\left.\frac{\Delta E}{\Delta t}\right|_{\rm QED},
\end{align}
which we evaluate numerically below.  

Second, to evaluate the mean-square transverse momentum transfer, we need $dN/d\tau d^2k_\perp$.  The derivation proceeds in parallel to the previous.  We use the change of variables described in the classical case \eq{dkzdtaureln}.  We keep the scaling constant $C$, this time determining its value by taking the classical limit with the result that $C=1$ (again).  Thus we obtain
\begin{align}\label{eq:dWd2kperpMsquared}
\frac{dW}{dt d^2k_\perp} &= \frac{1}{m}w(k_\perp^2,|eE|)
\end{align}
Then the mean-square transverse momentum transfer is simply
\begin{align}\label{eq:QEDkappa}
2\kappa_{\rm q}=\int \frac{dW}{dt d^2k_\perp} k_\perp^2 d^2k_\perp.
\end{align}
The classical limit commutes with the small $k_\perp$ limit, which could also be used to determine the scaling constant.  In the small $k_\perp$ region, $k_\perp^2\ll m^2,eE$, we find that QED predicts greater emission probability,
\begin{align}\label{eq:dNqoverdNcl}
\frac{dW/dtd^2k_\perp}{dN_\gamma^{\rm cl}/dtd^2k_\perp}=\frac{1}{1-e^{-\pi m^2c^3/eE\hbar}}\left(1+...\right),
\end{align}
which is a quantum effect (disappearing with $\hbar\to 0$) and only becomes significant for $eE\hbar/m^2c^3=a/m\sim 1$.  Similar to the Bose factor in the accelerated detector calculations \cite{Unruh:1976db}, it arises from the normalization of the wavefunctions which in turn is related to the hyperbolic functions in the classical particle action as recognized in analysis of spontaneous pair production \cite{PauchyHwang:2009rz,Labun:2012jf}.  

As we shall see in numerical evaluations of the differential emission rate, the phenomenology of photon emission does not change qualitatively with inclusion of electron recoil in QED.  As $a\sim m$, the rate of small $k_\perp$ emission is slightly enhanced \eq{dNqoverdNcl}.  For this reason--and ignoring the novel phenomena at strong fields $E\simeq m_e^2c^3/e\hbar$ especially pair creation--we argue that a Langevin equation should continue to model the electron-radiation dynamics.  We define the diffusion constant from the Langevin relation,
\begin{align}\label{eq:Dq}
D_{\rm q}=\frac{\kappa_{\rm q}\tau_{Dq}^2}{2m^2}.
\end{align}

\section{Comparison of Results and Discussion}

\begin{figure}
\includegraphics[width=0.5\textwidth]{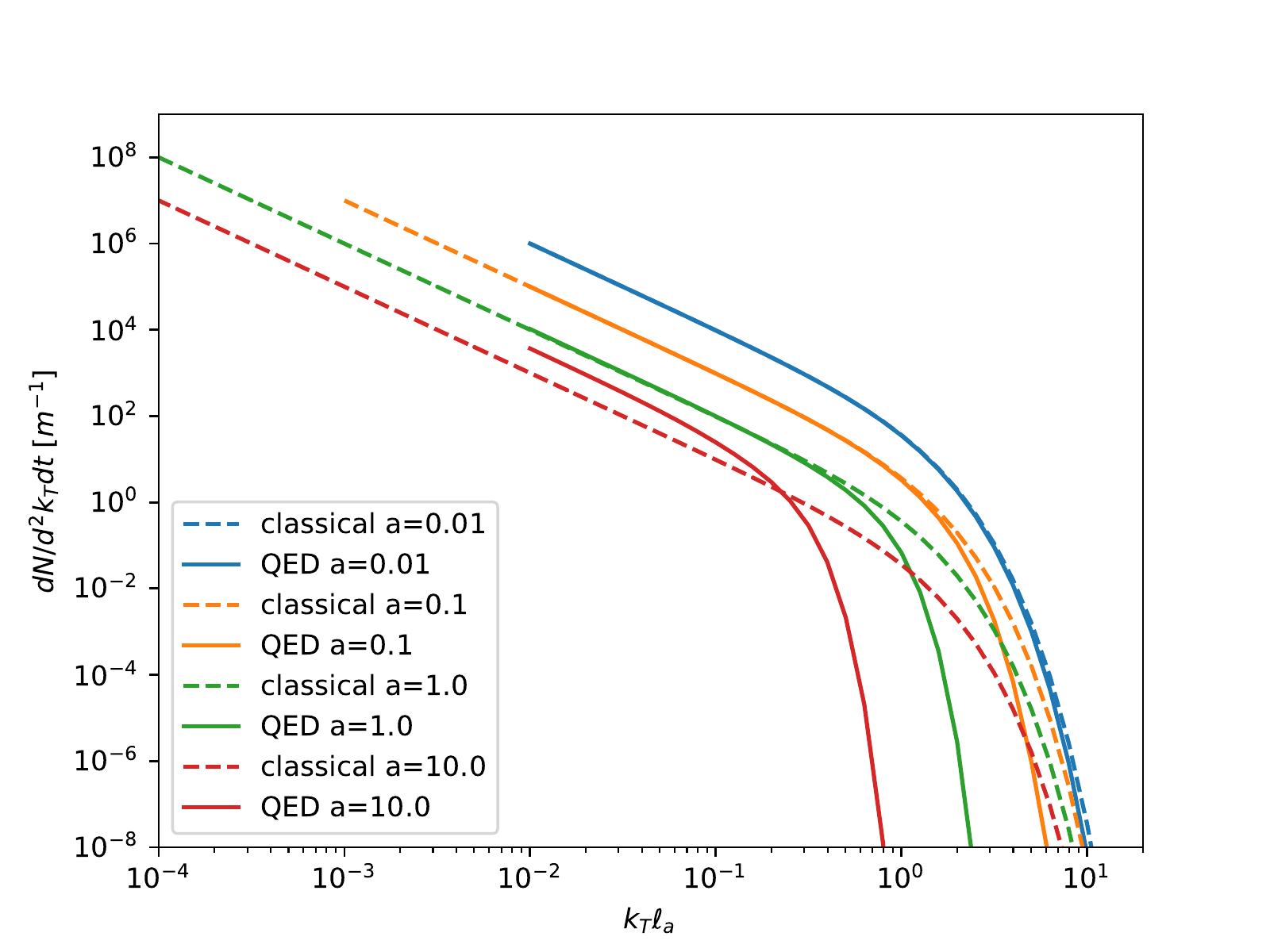}
\caption{\label{fig:dNdkT} The rate of photon emission per unit transverse momentum.  The wavenumber is normalized to the acceleration length scale $\ell_a=c^2/a=m/eE$, with curves comparing different magnitude of acceleration, normalized to $m$.  } %
\end{figure}

We now make quantitative comparisons of the observables computed in the previous section.  To establish intuition for the diffusion-related observables, we start with the photon emission rate differential in transverse momentum.  As shown in Figure \ref{fig:dNdkT}, the small $k_\perp$ behaviour is the same $dN/dtd^2k\perp\sim k_\perp^{-2}$ for classical and QED calculations, with the normalization of the QED result enhanced by the Bose-like factor \eq{dNqoverdNcl} visible for larger acceleration $a/m> 1$.  However for high $k_\perp\gtrsim 1/\ell_a$, QED predicts a significantly lower emission probability especially for $a/m\gtrsim 0.1$.  

\begin{figure}
\includegraphics[width=0.45\textwidth]{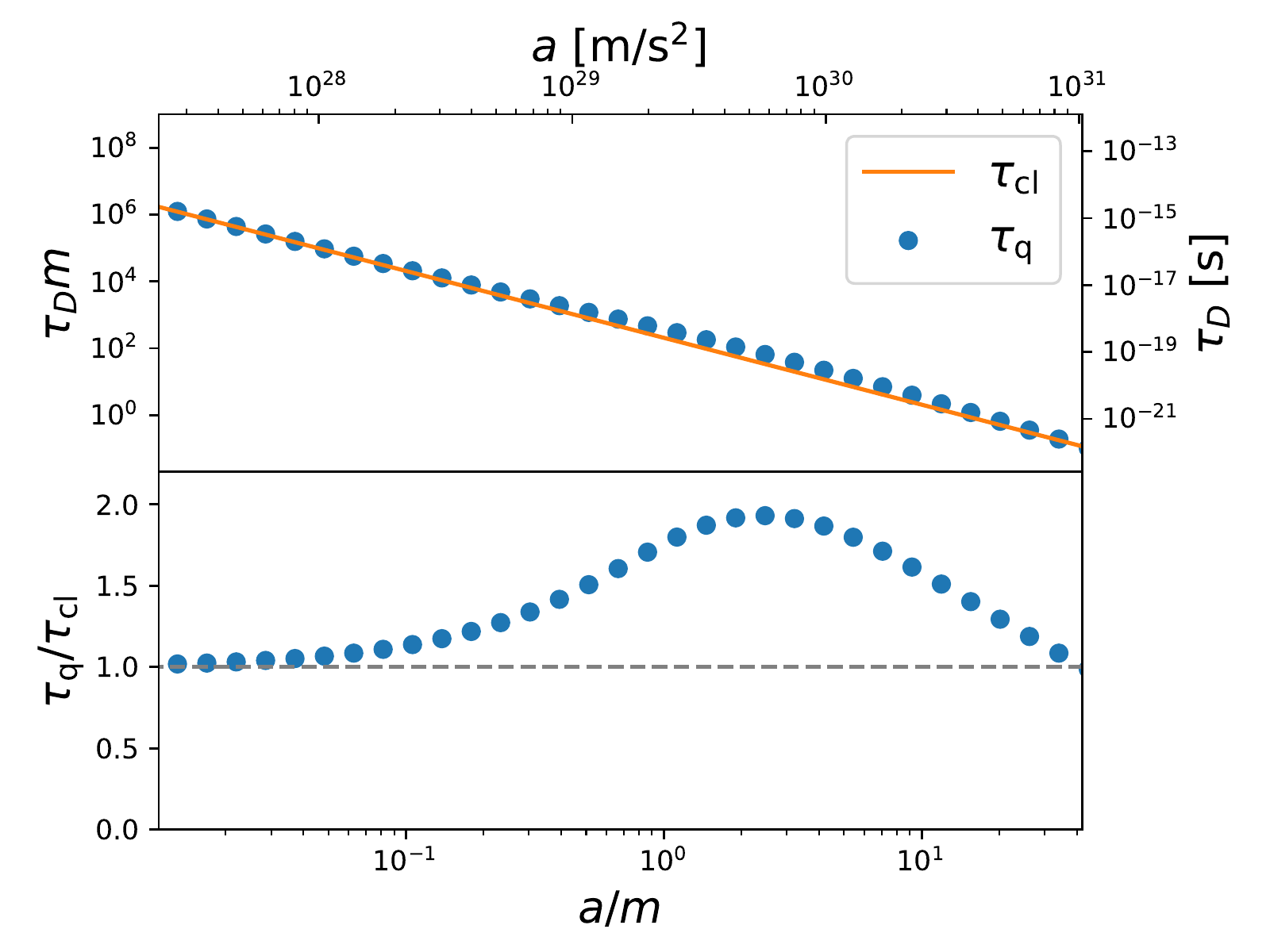}
\includegraphics[width=0.45\textwidth]{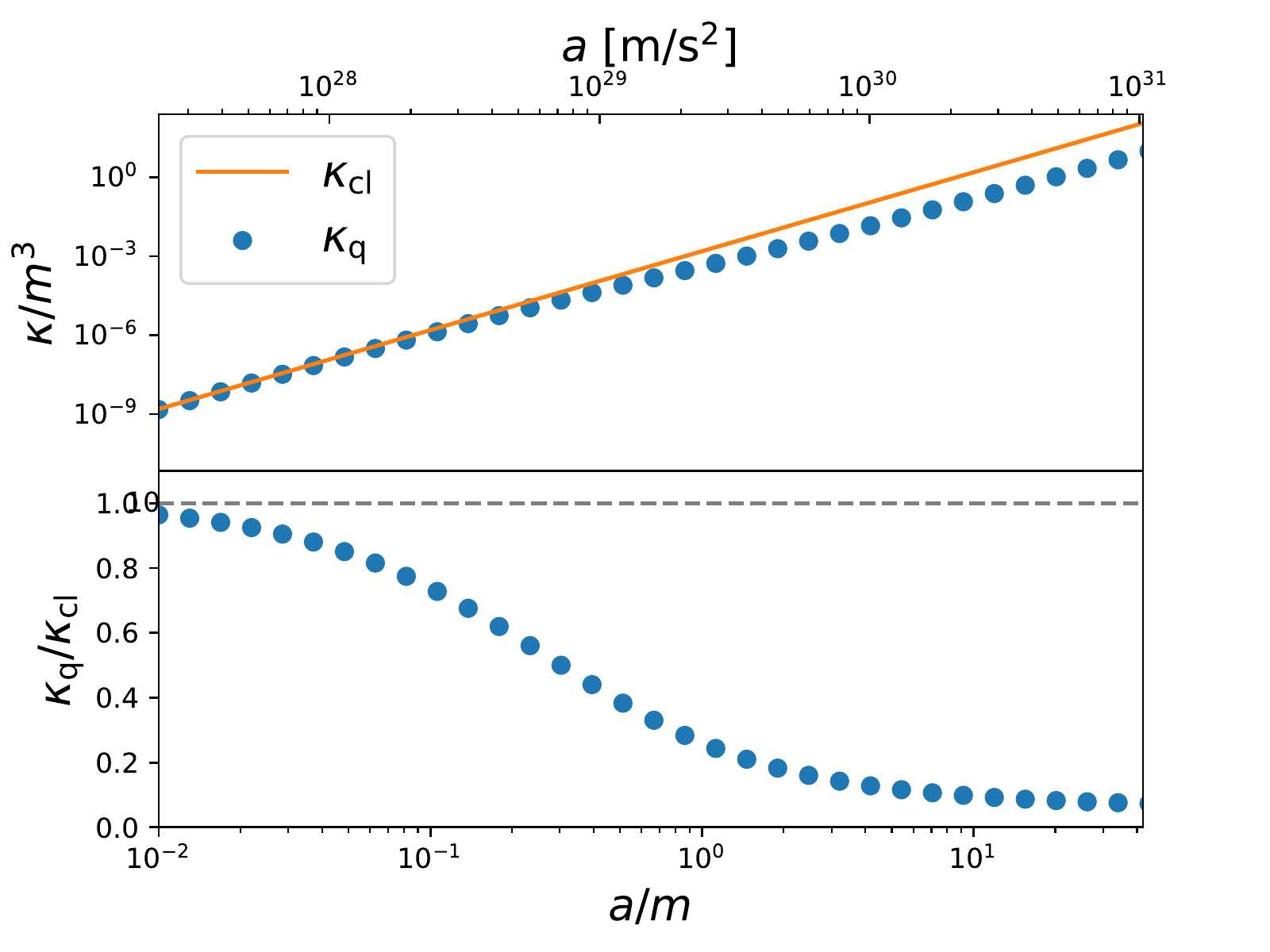}
\caption{\label{fig:taudkappa} Left: The dissipation time $\tau_D$ as a function of acceleration, classical radiation \eq{dampingtime} and QED \eq{QEDtauD} predictions.  Right: The mean-square momentum transfer to the electron obtained from classical \eq{kappaclresult} and QED \eq{QEDkappa}.}
\end{figure}

In classical calculations, the acceleration is the only variable scale and quantities such as the rate of energy loss and transverse momentum transfer should vanish as $a\to 0$.  The only other scale that can be involved is the LAD time scale $\tau_e$ \eq{taue}.  Considering first the damping time $\tau_D$ in Figure \ref{fig:taudkappa}, we find that QED predicts an enhancement from the classical result for $a/m<40$ and a suppression for $a/m\gtrsim 40$.  Since the differential emission rate \eq{dWd3kMsquared} is isotropic in transverse wavenumber, $d^2k_\perp=2\pi k_\perp dk_\perp$, and the resulting $k_\perp^2$ weight in the integrand cancels the $1/k_\perp^2$ divergence of the emission rate at small $k_\perp$.  This increases the importance of larger $k_\perp$ to the integral, where the QED differential probability is smaller, thus decreasing the energy loss rate.  The keen reader may notice small variations in the calculated value of $\tau_D$ around $a/m\simeq 0.1$ and later in $\kappa$ and derived quantities; these are numerical artifacts that seem to arise from challenges in finding a sufficiently accurate representation of the confluent hypergeometric functions in the differential QED emission rate.

In dimensionful units, the damping time is of order 1 femtosecond for an acceleration $a/m\simeq 0.01$ corresponding to an electric field $|E|\simeq 10^{16}$\,V/m.  As observed in Ref. \cite{Iso:2010yq}, this is the timescale and therefore the electric field strength that would be required if thermalization were desired within a single cycle of a laser pulse, as proposed by Ref. \cite{Chen:1998kp}.  However, more recent calculations for oscillating trajectories show that a model detector does not converge to equilibrium at the temperature $T_a$ \cite{Doukas:2013noa}.  Laser wakefield acceleration utilizes (co-moving) quasi-stationary longitudinal electric fields, which persist over $\sim$10 cm of propagation or 0.3 ns.  If we require thermalization within half of that acceleration time (150 picoseconds), the electric field should be $|E|\simeq 2.4\times 10^{13}$\,V/m.  The longitudinal fields generated during laser wakefield acceleration $\sim 10^{11}$ V/m remain orders of magnitude lower.  Conversely, for $|\vec E|\simeq 10^{11}$ V/m, the acceleration would have to persist for $\sim 10$ microseconds to exceed the dissipation time, corresponding to an acceleration length of 3\,km.  Conventional radio-frequency accelerators that are actually 3\,km long fare worse, with maximum accelerating gradients of $\sim 10^8$ V/m, which due to the $a^{-2}$ scaling of $\tau_D$ would require an acceleration time of 10 seconds or length of $3\times 10^6$\,km.  This estimate obviously assumes that focusing elements interspersed between $\sim 1$-2\,m accelerator chambers do not interfere with considering the acceleration approximately constant, and every accelerator chamber provides the same accelerating gradient.

In the classical limit, the mean-square momentum transfer per unit time is a function of only $a$.  In the comparison to QED, the $k_\perp^3$ weight in the integrand ensures that the high-$k_\perp$ region is still more important in determining the integral and the QED result $\kappa_{\rm q}$ is less than the classical result $\kappa_{\rm cl}$ for all values of $a$.  

Aside from the dissipation time setting the scale for the required duration of the acceleration, the diffusion constant is next most important step toward a measurement.  For a heavy particle in a thermal bath, the diffusion constant describes the linear growth of the mean square displacement in time.  In the present dynamics, it describes the linear growth of the transverse size of a hypothetical electron beam being accelerated.  However in accelerator physics the mean square displacement alone is typically not measured, and the calculation here should be consider a stepping-stone to more specialized observables.

\begin{figure}
\includegraphics[width=0.5\textwidth]{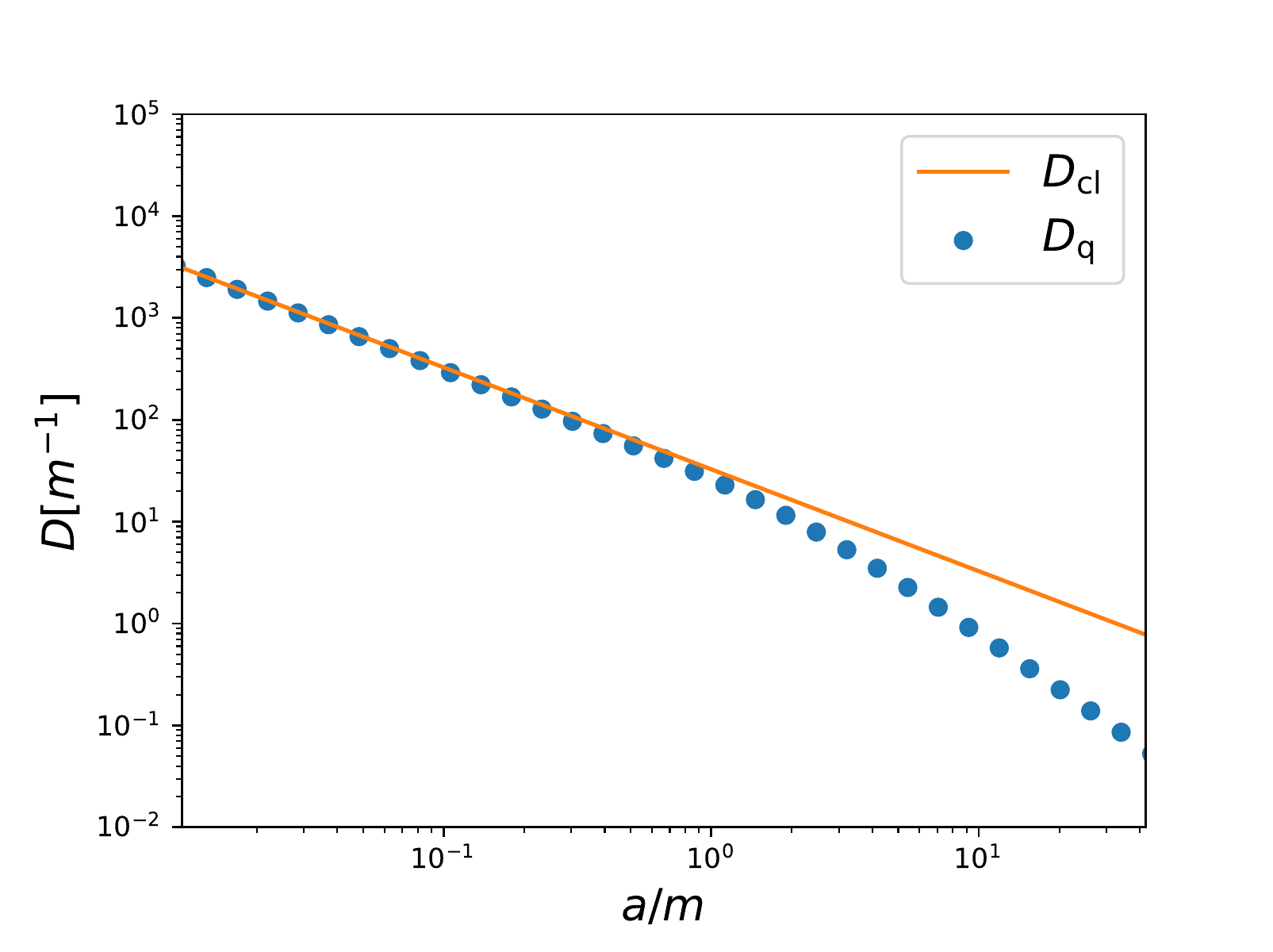}
\caption{\label{fig:diffusion} Diffusion constant derived from classical \eq{Dcl} and QED \eq{Dq} radiation dynamics.  }
\end{figure}

The diffusion constant is a combination of $\tau_D$ and $\kappa$, and since $\tau_D\propto a^{-2}$ and $\kappa\propto a^3$ the diffusion constant $D\sim a^{-1}=T^{-1}$.  This inverse proportionality contrasts with diffusion associated with nonrelativistic Brownian motion but is typical for diffusion in massless gauge theories.  An intuitive reason for this inverse proportionality is that, as massless particles, the number density of photons increases with temperature.  Therefore the density of scatterers rises with temperature and increases the rate of soft, largely dissipative scattering events.  This picture is consistent with the finding that QED further enhances the emission rate at small $k_\perp$ and results in a smaller diffusion constant, shown in Figure \ref{fig:diffusion}.  

However electron diffusion in a low temperature ($T\ll m_e$) QED plasma or heavy quark diffusion in a QCD plasma ($\Lambda_{QCD}\ll T\ll m_Q$) differ from the results for constant acceleration in their manifest dependence on the coupling constant $e^2$.  Statistical definitions of the dissipation time and mean-square momentum transfer involve squared matrix elements (as they did implicitly in \ssec{quantphoton} \ssec{QED}), schematically \cite{Svetitsky:1987gq,Braaten:1991jj,Moore:2004tg}
\begin{align}
\frac{1}{\tau_D}&=\frac{1}{|\vec v|}\frac{dE}{dt}=\int [dk][dk'][dp'](p'_0-p_0)|\mathcal{M}|^2n_b(\vec k_\perp)\big(1+n_b(\vec k'_\perp)\big)\\
N_d\kappa&=\frac{1}{2m}\int [dk][dk'][dp'](\vec p'_\perp-\vec p_\perp)^2|\mathcal{M}|^2n_b(\vec k_\perp)\big(1+n_b(\vec k'_\perp)\big)
\end{align}
where the phase space integrals $[dk]\equiv d^3k/(2\pi)^3$ come also with momentum conserving $\delta$ functions.  The matrix elements are $2\to 2$ scattering amplitudes, e.g. linear Compton scattering for an electron in a QED plasma.  The phase space integrals therefore involve an incoming photon momentum $k$ and outgoing photon momentum $k'$, each matrix element is proportional to $e^2$, and the observables $\tau_D^{-1},\kappa$ are proportional to $\alpha^2$.  In \eq{Dcl}, one power of $e$ is hidden in the acceleration, $D\propto (e^2a)^{-1}\sim (e^3E)^{-1}$, and one might argue that the missing power of $e$ would be restored on considering the source of the $\vec E$ field from Maxwell's equation $\partial_\mu F^{\mu\nu}=j^\nu\sim enu^\nu$.

Last, we plot the product of the damping time and mean-square momentum transfer, $\tau_D\kappa/2mT_a$.  In the classical limit, this combination is a constant equal to 1 \eq{kappacltauT}.  Combining the QED results, we find that the ratio is suppressed from the classical value for all values of $a$, approaching zero for $a\gg m$.  This combination of observables, related by the Langevin dynamics to the mean-squared transverse momentum in equilibrium $\langle p_\perp^2\rangle$, shows the fastest deviation from the classical result as $a$ increases.

\begin{figure}
\includegraphics[width=0.5\textwidth]{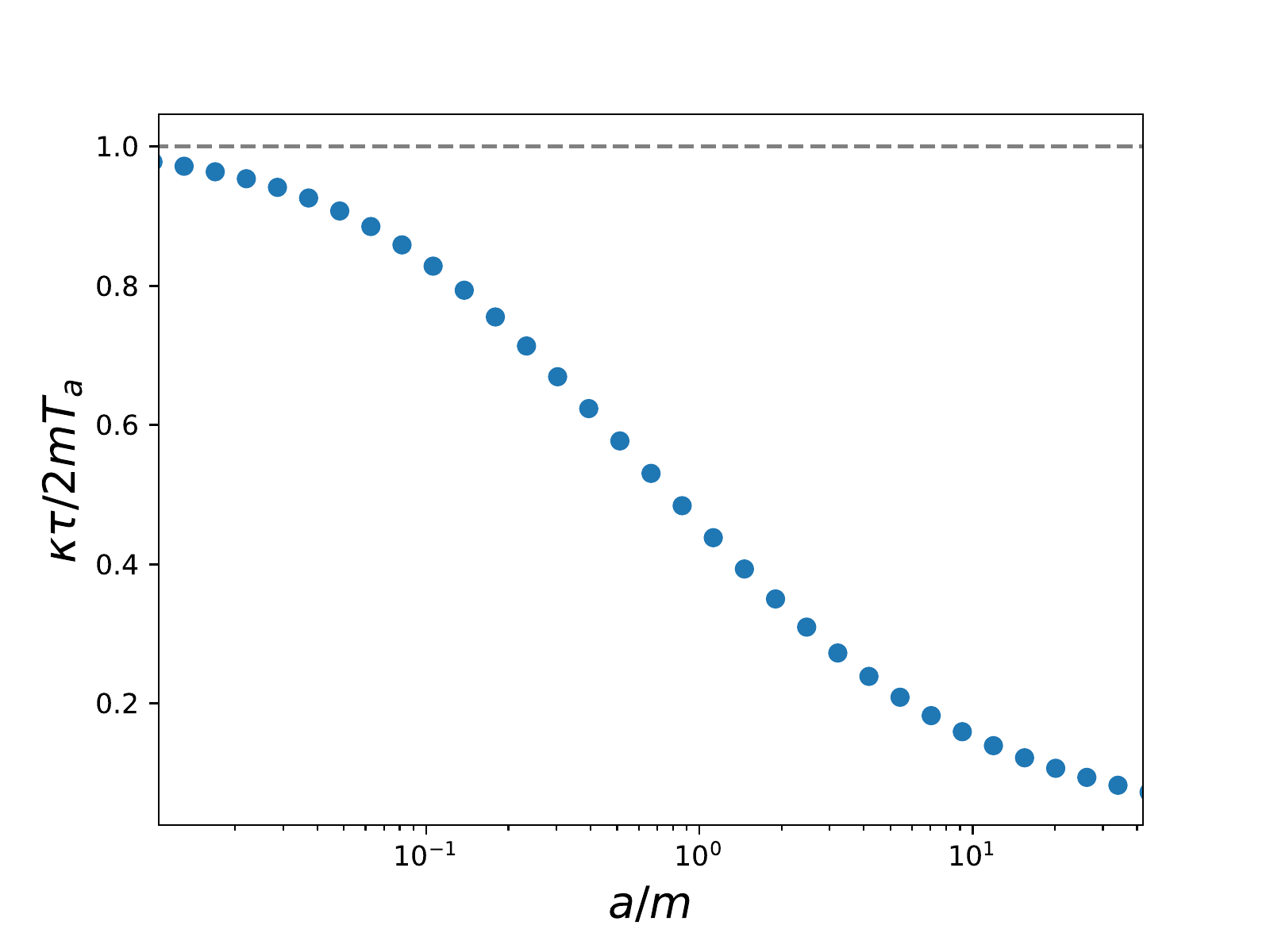}
\caption{\label{fig:equipartition} The product $\kappa_q\tau_q$ normalized to its classical value $2mT_a$.  }
\end{figure}

The mean-square transverse momentum Figure \ref{fig:equipartition} or the transverse diffusion Figure \ref{fig:diffusion} likely provide the most useful observables to study experimentally.  Though we have found quite small QED corrections, we could with sufficient statistics and precise control at least verify the classical radiation predictions.  An experiment based on laser wakefield acceleration requires substantial improvements in the control and consistency of the acceleration dynamics to be successful.  Transverse momentum oscillations, which can approach $|p_\perp|\sim m$ in magnitude, will have to be accounted for, though it is possible that radiation reaction \eq{LAD} gradually suppresses the oscillations in the absence of a driving force.

The description here of particle dynamics in strong-field QED regime is of course incomplete.  The characteristic timescale for the dissipation of field energy into electron-positron pairs is exponential in the electric field strength, with a field providing an acceleration $a\gtrsim 0.2m$ decaying on the order of 3 ps \cite{Labun:2009vdt}.  Higher order in $\alpha$ processes, such as the direct bremsstrahlung of a pair by the electron are not likely to be important until $a\sim m$.  These dynamics are expected to correct the calculations here in the $a\gtrsim m$ regime.

In summary, we have found that thermalization of a probe particle (electron) undergoing constant acceleration is due to its classical radiation.  Nonzero variance in the mean-square transverse momentum (chosen for being invariant under boosts compatible with the symmetry of constant acceleration) is explained by computing the second momentum of the radiation distribution, and $\hbar$ only enters as a matter of converting units of photon wavenumber to electron momentum.  We expect that the diffusion-related observables obtained here by way of the classical photon number can also be obtained from the appropriate correlator of the classical radiation field, similar to QED and QCD calculations \cite{Casalderrey-Solana:2006fio}.  Such a calculation would be interesting in revealing how $\hbar$ enters.  Building on the work of Refs. \cite{Galley:2005tj,Iso:2010yq}, our discussion emphasizes the origin of the characteristic features of a thermal system in the model of the radiation dynamics.  Specifically, the uncorrelated nature of the noise is valid in the classical regime where most emission is soft and dissipative while rarer hard emissions drive the momentum fluctuations.  It follows that any more nuanced description of the radiation dynamics, e.g. bringing in higher order correlations from the trajectory, will generally break the perfectly thermal relations obtained here.  The quantitative results give an idea of the experimental challenge in observing effects of the acceleration temperature.  Laser wakefield accelerators provide the best combination of field strength and acceleration length, but are still a factor $\sim 100$ too weak field or too short duration.  Although some increase of both may be possible in wakefield accelerators e.g. by using ``flying focus'' laser wakefield schemes or a combination of laser and beam-driven wakefields, these numbers suggest that we will require more precise calculations of well-defined electron beam observables and high-statistics measurements to distinguish the impact of this ``thermalization'' effect for constant acceleration.

\begin{acknowledgments}
This material is based upon work supported by the National Science Foundation under Grant PHY-2108921 and the Air Force Office of Scientific Research under Grant FA9550-17-1-0264. BMH and LL acknowledge support from Tau Systems, Inc.  GT acknowledges financial support from FAPESP grant 2021/01700-2 and Bolsa de produtividade CNPQ 306152/2020-7.  HT acknowledges support from the Conselho Nacional de Desenvolvimento Cient\'{i}fico e
Tecnol\'{o}gico (CNPq), proc. 141024/2017-8.
\end{acknowledgments}


\begin{appendix}



\section{Transverse photon emission rate: classical calculation}\label{app:dNclcalc}

The calculation of the photon emission rate is available from many references \cite{nikishov1985problems,Higuchi:1992both,Biro:2011ne}, so we here just highlight the small refinements in our derivations with respect to present goals.  For an electron in a constant electric field $\vec E=|\vec E|\hat z$, the 4-velocity $u^\mu$ and trajectory $\xi^\mu$ is equivalent to that under constant acceleration,
\begin{align}\label{eq:cltraj}
u^{\mu}&=\big(\cosh(a\tau/c),u_x(0),u_y(0),c\sinh(a\tau/c)\big)\\
\xi^{\mu}(\tau)&=\big((c/a)\sinh(a\tau/c),u_x(0)\tau,u_y(0)\tau,(c^2/a)\cosh(a\tau/c)\big).
\end{align}
For notational simplicity we continue with the electron $p_\perp=0$ case.  We start from the classical formula for the emitted photon number \cite{Jackson:1998nia}
\begin{align}\label{eq:Pclgeneral}
dN^{\rm cl}_{\gamma}=\frac{e^2}{8\pi^2 c|\vec k|^2}|\vec A(\vec k)|^2d^3k
\end{align} 
with the Fourier transformed vector potential determined by the Lienard-Wiechert potentials,
\begin{align}\label{eq:Aclgeneral}
A(\vec k)=\int e^{i\varphi(t)}\frac{d}{dt}\left[\frac{\vec n\times(\vec n\times\vec \beta)}{1-\vec n\cdot\vec\beta}\right]dt \\ \notag
\varphi(t)=k(ct-\vec n\cdot\vec\xi), \qquad c\vec\beta=\frac{d\vec\xi}{dt},
\end{align}
where $\vec\beta=\vec u/u^0$ is the normalized 3-velocity of the electron, $\vec n$ is the unit vector in the direction of the emission and $k=|\vec k|$ is the magnitude of the wave vector.
It is convenient to change variables to electron rapidity $y$, linearly related to the proper time $\tau$, and photon rapidity $\eta$, related to the angle of emission,
\begin{align}
\tanh y&=|\vec\beta|=\tanh(a\tau/c),\\
\tanh\eta&=\vec n\cdot\vec\beta=k_z/k
\end{align}
so that the phase factor in \eq{Aclgeneral} reads simply
\begin{align}
\varphi(y)=\frac{c^2k_\perp}{a}\sinh(y-\eta).
\end{align}
Changing integration variables $dt\to dy$, the vector cross product in the integrand is written in terms of (constant) transverse polarization vectors on the unit sphere $\vec\epsilon_{\Omega}$, such that vector in square brackets in \eq{Aclgeneral} is
\begin{align}
\frac{d}{dy}\left[\frac{\vec n\times(\vec n\times\vec \beta)}{1-\vec n\cdot\vec\beta}\right]=\vec\epsilon_{\Omega}\frac{\cosh\eta}{\cosh^2(y-\eta)}.
\end{align}
Rather than using these two expressions to evaluate the integral in \eq{Aclgeneral}, we write out the squared vector potential in \eq{Pclgeneral},
\begin{align}
dN^{\rm cl}_{\gamma}=\frac{e^2}{4\pi^2 c|\vec k|^2}d^3k\int dydy'\frac{\exp\left(i(k_\perp/a)(\sinh(y-\eta)-\sinh(y'-\eta)\right)}{\cosh^2(y-\eta)\cosh^2(y'-\eta)}
\end{align}
and change variables to average $2\bar y=y+y'$ and relative rapidity $r=y-y'$.  After some algebra, the integrand depends only $\bar y-\eta$,
\begin{align}
dN^{\rm cl}_{\gamma}=\frac{e^2}{4\pi^2 c|\vec k|^2}d^2k_\perp dk_z\int d\bar ydr\frac{\exp\left(2i(k_\perp/a)\sinh(r/2)\cosh(\bar y-\eta)\right)}{(\cosh(2(\bar y-\eta))+\cosh(r))^2},
\end{align}
where $k_z=k_\perp\sinh\eta$.
Changing the integration variable for the photon longitudinal wavenumber to the photon rapidity, $ dk_z=ck_\perp\cosh\eta d\eta$, we integrate over $\eta$ first, shifting $\eta\to\bar y-\eta$ with no change to the integrand since the integration domain is $(-\infty,\infty)$.  Having eliminated dependence on $\bar y$, we undo much of the algebra and change variables $(r,\eta)\mapsto (z=\eta+r/2,z'=\eta-r/2)$ to obtain two decoupled complex conjugate integrals.  The result is
\begin{align}
dN^{\rm cl}_{\gamma}=\frac{e^2}{4\pi^2 }d^2k_\perp d\bar y\left|\int_{-\infty}^{\infty} dz\frac{\exp\left(i(k_\perp/a)\sinh z\right)}{\cosh^2(z)}\right|^2
\end{align}
The integral then yields the modified Bessel function $K_0'(k_\perp/a)=-K_1(k_\perp/a)$.

The mean square momentum transfer integral is made dimensionless by scaling $k_\perp=|\vec k_\perp|\to k_\perp/a$,
\begin{align}
2\kappa_{\rm cl}=\frac{1}{\hbar}\int d^2k_\perp (\hbar k_\perp)^2 \frac{dN^{\rm cl}_\gamma}{d\tau d^2k_\perp}
=\frac{2\alpha}{\pi}a^3\frac{\hbar^2}{c}\int_0^\infty x^3|K_1(x)|^2dx
\end{align}
and evaluated using Eq. 6.576 of Ref. \cite{gradshteyn2014table},
\begin{align}\label{eq:Besselintegralid}
\int_0^\infty x^{-\lambda}K_\mu(ax)K_\nu(bx)dx=&\frac{2^{-2-\lambda}a^{-\nu+\lambda-1}b^{\nu}}{\Gamma(1-\lambda)}\Gamma\left(\frac{1-\lambda+\mu+\nu}{2}\right)\Gamma\left(\frac{1-\lambda-\mu+\nu}{2}\right) \notag \\
&\times\Gamma\left(\frac{1-\lambda+\mu-\nu}{2}\right)\Gamma\left(\frac{1-\lambda-\mu-\nu}{2}\right)
\notag \\
&\times \phantom{l}_2F_1\left(\frac{1-\lambda+\mu+\nu}{2},\frac{1-\lambda-\mu+\nu}{2};1-\lambda;1-\frac{b^2}{a^2}\right) \\ \notag
&\mathrm{Re}(a+b)>0 \qquad \mathrm{Re}\lambda<1-|\mathrm{Re}\mu|-|\mathrm{Re}\nu|
\end{align}
Since the $a=b=1$ in our case, the confluent hypergeometric function is evaluated at 0, which for all values of the parameters reduces to 1.  The product of $\Gamma(z)$ functions and $2^{-1}$ reduces to the constant $2/3$, to arrive at the result quoted in the text \eq{kappaclresult}.

\section{Transverse photon emission rate: QED calculation}\label{app:sfQED}

We wish to compute the emitted photon distribution fully differential in photon momentum,
\be\label{eq:dWd3kMsquaredapp}
dW=\frac{d^3k}{(2\pi)^32k_0}\frac{1}{2}\sum_{\sigma,\sigma'}\sum_{\epsilon,\epsilon'}\int\frac{d^3p'}{(2\pi)^32E_{p'}}\left|\mathcal{M}[e_{\vec p}\to e_{\vec p'}\gamma_{\vec k}]\right|^2,
\ee
summed over final electron spin and photon polarization and averaged over initial electron spin.
The matrix element is
\be
-i\mathcal{M}[e_p\to e_{p'}\gamma_k]=-ie\int d^4x \bar\psi_{\sigma',p'}^{(+)}(x)\slashed{\epsilon}^*\frac{e^{ikx}}{\sqrt{2k_0}}\psi_{\sigma,p(+)}(x).
\ee
where $\psi_{\sigma,p(+)}(x)$ is the incoming electron wavefunction and $\bar\psi_{\sigma',p'}^{(+)}(x)$ is the outgoing electron wavefunction.  The wavefunctions are solutions to the Dirac equation with a classical external vector potential corresponding to an electric field in the $\hat z$ direction,
\begin{align}\label{eq:DEAcl}
\left(i\slashed{\partial}_x-e\slashed{A}_{\rm cl}(x)-m\right)\psi(x)=0, \qquad A^{\mu}_{\rm cl}(x)=\delta^\mu_3Et.
\end{align}
Going to the second order equation with the Ansatz $\psi(x)=\left(i\slashed{\partial}-e\slashed{A}_{\rm cl}+m\right)\psi^{(2)}(x)$ and changing variables to $u=\sqrt{2/eE}(p_z-eEt)$ leads the parabolic cylinder differential equation
\be
\left(\partial_u^2+\lambda\pm\frac{i}{2}+\frac{u^2}{4}\right)f_\lambda(u)=0.
\ee
The complete set of solutions is $D_{i\lambda}(-e^{-i\pi/4}u), D_{i\lambda-1}(e^{-i\pi/4}u),D_{-i\lambda}(-e^{i\pi/4}u),D_{-i\lambda-1}(e^{i\pi/4}u)$.  A detailed derivation of the wavefunctions with updated notation in Ref. \cite{Anderson:2013both} and the results are \cite{nikishov1985problems}
\begin{align}
\psi_{\sigma\lambda(\pm)}(x)&=N_\lambda\sqrt{2eE}e^{-\pi\lambda/4\pm i\zeta_\lambda}e^{i\vec p\cdot\vec x}\chi_{\sigma\lambda(\pm)}(u), \\
\sqrt{2eE}\chi_{\lambda,1(+)}(x)&=e^{i\pi/4}(i\lambda)D_{i\lambda-1}(-\xi)u_2+\frac{p_1u_3\!+\!(m\!-\! ip_2)u_1}{\sqrt{2eE}}D_{i\lambda}(-\xi), \\
\sqrt{2eE}\chi_{\lambda,2(+)}(x)&=-e^{i\pi/4}D_{i\lambda}(-\xi)u_1+\frac{p_1u_4+(m+ip_2)u_2}{\sqrt{2eE}}D_{i\lambda-1}(-\xi),\\
\sqrt{2eE}\chi_{\lambda,1(-)}(x)&=e^{-i\pi/4}D_{-i\lambda}(-\xi^*)u_2+\frac{p_1u_3\!+\!(m\!-\! ip_2)u_1}{\sqrt{2eE}}D_{-i\lambda-1}(-\xi^*), \\
\sqrt{2eE}\chi_{\lambda,2(-)}(x)&=e^{-i\pi/4}i\lambda D_{-i\lambda-1}(-\xi^*)u_1+\frac{p_1u_4+(m+ip_2)u_2}{\sqrt{2eE}}D_{-i\lambda}(-\xi^*).
\end{align}
where we have defined for notational simplicity, $\xi=e^{-i\pi/4}u$, $\zeta_\lambda=(\lambda/2)(1-\ln\lambda)$ and an orthogonal and complete basis of spinors,
\be\label{eq:Efieldspinors}
u_1=\begin{pmatrix}1\\0\\0\\1 \end{pmatrix}, \quad
u_2=\begin{pmatrix}0\\1\\1\\0 \end{pmatrix}, \quad
u_3=\begin{pmatrix}1\\0\\0\\-1 \end{pmatrix},\quad
u_4=\begin{pmatrix}0\\1\\-1\\0 \end{pmatrix}.
\ee
Using that the outgoing electron solution is equivalent to the time-reversed incoming positron solution, $\psi^{(+)}(t,\vec x)=\psi_{(-)}(-t,\vec x)$, we have
\begin{align}\label{eq:Msimplified}
-i\mathcal{M}&=\frac{-ie}{\sqrt{2k_0}}N_\lambda N_{\lambda'}^* (2eE)e^{-(\lambda+\lambda')\pi/4+i(\zeta_{\lambda'}+\zeta_\lambda)}\int d^4x e^{ikx}e^{i(\vec p'-\vec p)\cdot\vec x}\chi_{\sigma'\lambda'(-)}^{\dag}(-u)\gamma^0\slashed{\epsilon}^*\chi_{\sigma,\lambda(+)}(u) 
\end{align}
The spatial integrals can be done immediately to yield 3-momentum conservation $\vec {p'}=\vec p-\vec k$.  Integrating over the final state momentum with the $\delta$ function, and after extensive algebra to reduce the remaining $t$ integral, the fully differential rate is \cite{nikishov1985problems}
\begin{align}\label{eq:finalavgdWd3kphoton}
dW&=\frac{d^3k}{(2\pi)^32k_0}\int \frac{d^3p'}{(2\pi)^32p_0'}\frac{1}{2}\sum_{\epsilon,\epsilon',\sigma,\sigma'}\!\!|\mathcal{M}|^2 \notag \\
&=\frac{d^3k}{(2\pi)^32k_0}\mathcal{N}\pi e^{-\frac{3\pi}{4}\frac{k_\perp^2-2\pTkT}{eE}}\left\{\left(2E_\perp^2+k_\perp^2-2\pTkT\right)\frac{k_\perp^2}{E_\perp^2}|\Psi'|^2
\right. \\ \notag 
&\left. +(2p_\perp^2+k_\perp^2-2\pTkT)|\Psi|^2 -\left(\frac{2p_\perp^2k_\perp^2}{E_\perp^2}+\frac{2(2\pTkT-k_\perp^2)}{E_\perp^2}(E_\perp^2-\pTkT)\right)\mathrm{Re}[\Psi'\Psi^*]\right\} 
\end{align}
where $E_\perp^2=p_\perp^2+m^2$.  The wavefunction normalizations have been combined into 
\begin{align}
\mathcal{N}&=\frac{2e^2\exp\big(-\pi(\lambda+\lambda')/2)\big)}{2\lambda'(eE)^2(1-e^{-2\pi\lambda})(1-e^{-2\pi\lambda'})},
\end{align}
 and $\Psi$ is the confluent hypergeometric of the second kind, evaluated at
\begin{align}
\Psi &\equiv \Psi\!\left(\frac{iE_\perp^2}{2eE},1-\frac{i(k_\perp^2-2\pTkT)}{2eE};\frac{-ik_\perp^2}{2eE}\right),
\end{align}
which is related to the confluent hypergeometric$~_1F_1(a,b;z)$ by
\begin{align}\label{eq:hypergeomreln}
\Psi(a,b;z)=\frac{\Gamma(1-b)}{\Gamma(a-b+1)}\,_1F_1(a,b;z)+\frac{\Gamma(b-1)}{\Gamma(a)}z^{1-b}\,_1F_1(a-b+1,2-b;z)
\end{align}
The $\hbar\to 0$ limit yields the classical result \cite{nikishov1985problems}.

\section{Results for a scalar radiation field}

The number of particles emitted by a classical source \( J(x) \) on a general scalar field \( \varphi \) can be found in standard textbooks \cite[Chapter 2]{Peskin:1995ev}, and is given by
\begin{equation}
  \int dN  =  \int \frac{d^{3}p}{(2\pi)^{3}} \frac{1}{2E_{p}} \abs{J(p)}^{2}.
\end{equation}
For a classical charged particle source following an accelerated trajectory \eq{cltraj}, we have
\begin{equation}
  J(x; \xi)  =  e \int d\tau \sqrt{u^{2}(\tau)} \delta^{4}(x - \xi(\tau)),
\end{equation}
and the Fourier transform of the source for the localized particle is given by
\begin{equation}  \label{eq:FourierTransformScalarJ}
  J(p)  =  e \int d\tau\, \exp( i (E_{p} / a) \sinh a\tau - i (p_{z} / a) \cosh a\tau ).
\end{equation}
We are interested in the number of photons emitted per unit transverse momentum and per unit proper-time \( dN/d\tau d^{2}p_{\perp} \), which can be obtained from the evaluation of the differential in 3-momentum
\begin{equation} \label{eq:dN/d3p}
  \frac{dN}{d^{3}p}  = \frac{1}{(2\pi)^{3} 2 E_{p}} \abs{J(p)}^{2}.
\end{equation}
Changing into relative coordinates \( \bar{\tau} \equiv \frac{1}{2} (\tau + \tau^{\prime}) \) and \( \delta \tau \equiv \frac{1}{2} (\tau - \tau^{\prime}) \), we can write the expression for the square of the current's Fourier transform as
\begin{equation}
  \abs{J(p)}^{2}  =  e^{2}\int d\tau d\tau^{\prime} \exp \left[(2i/a) \sinh a \delta\tau (E_{p}\cosh a\bar{\tau}- p_{z} \sinh a\bar{\tau})\right].
\end{equation}
Now we parametrize the particle's momentum by new hyperbolic variables \( E_{p} = p_{\perp} \cosh \eta \), \( p_{z} = p_{\perp} \sinh \eta \), and obtain the alternative representation
\begin{equation}
  \abs{J(p)}^{2}  =  2e^{2} \int d\delta\tau d\bar{\tau} \exp\left[(2ip_{\perp} / a) \sinh(a \delta\tau) \cosh(\eta - a \bar{\tau})\right],
\end{equation}
which makes clear that the integral is independent of the rapidity \( \eta \).  We can remove \( \eta \) directly from the integral, which would yield an exact expression \cite[Eq. 8.432-5]{gradshteyn2014table} for \( \abs{J(p)}^{2} \) as
\begin{equation}
  \abs{J(p)}^{2} = e^{2}\abs{\int d\tau \exp \left(i(p_{\perp} / a) \sinh a\tau\right)}^{2} = \frac{4e^{2}}{a^{2}} K_{0}^{2}(p_{\perp} / a).
\end{equation}
where \( K_{0} \) is a modified Bessel functions of the second kind.  In terms of the \( \eta \) coordinate, the \( dN \) differential takes the form
\begin{equation}
  \frac{dN}{d\eta\, d^{2}p_{\perp}} = \frac{1}{2 (2\pi)^{3}} \abs{J(p_{\perp})}^{2},
\end{equation}
and the final expression is given by
\begin{equation} \frac{dN}{d\eta\, d^{2}p_{\perp}} = \frac{e^{2}}{4\pi^{3} a^{2}}K_{0}^{2}(p_{\perp} / a).
\end{equation}
We can also obtain the same espression in terms of a differential on the mean proper-time \( \bar{\tau} \).  First we integrate over all longitudinal momenta
\begin{equation}
  \int dp_{z} \frac{dN}{d^{3} p} = \frac{1}{(2\pi)^{3}}\int \frac{dp_{z}}{2E_{p}} \abs{J(p)}^{2}.
\end{equation}
In terms of the momentum rapidity \( \eta \), we get
\begin{equation}
  \frac{dN}{d^{2} p_{\perp}}
  =
  \frac{e^{2}}{(2\pi)^{3}}
  \int d\eta\, d\delta\tau d\bar{\tau}
  \exp \left[
    (2 i p_{\perp} / a) \sinh(a \delta\tau) \cosh(\eta - a \bar{\tau})
  \right].
\end{equation}
Changing variables for the \( \eta \)-integral and extracting the linearly divergent total proper-time \( \int d\bar{\tau} \), we get
\begin{equation}
  \frac{dN}{d \bar{\tau} d^{2} p_{\perp}} 
  =
  \frac{e^{2}}{(2\pi)^{3}}
  \int d\eta\, d\delta\tau
  \exp \left[
    (2 i p_{\perp} / a) \sinh(a \delta\tau) \cosh(\eta)
  \right].
\end{equation}
Again from Eq. 8.432-5 of Ref. \cite{gradshteyn2014table}, we get an exact expression for the integral in terms of another modified Bessel function of the second kind
\begin{equation}
  \int d\delta\tau
  \exp \left[
    (2 i p_{\perp} / a) \sinh(a \delta\tau) \cosh(\eta)
  \right]
  =
  \frac{2}{a} K_{0}(2 (p_{\perp} / a) \cosh \eta),
\end{equation}
and the remaining \( \eta \) integral can be evaluated to (cf. Eq. 6.663-1 of \cite{gradshteyn2014table})
\begin{equation}
  \int d\eta\, K_{0}(2 (p_{\perp} / a) \cosh \eta)
  =
  K_{0}^{2}(p_{\perp} / a).
\end{equation}
The final result for the distribution of scalar particles created per transverse momentum and proper time is thus
\begin{equation}
  \frac{dN}{d \bar{\tau} d^{2} p_{\perp}}
  =
  \frac{e^{2}}{4\pi^{3} a}
  K_{0}^{2}(p_{\perp} / a),
\end{equation}
which coincides with the previous direct calculation from the ``momentum rapidity'' \( \eta \) by the direct substitution \( \eta \leftrightarrow a\bar{\tau} \).

We are interested in the mean squared transverse momentum transfer for the theory, so we calculate
\begin{equation}
  2 \kappa_{\text{scalar}}
  =
  \int d^{2} p_{\perp} p_{\perp}^{2} \frac{dN}{d\bar{\tau} d^{2}p_{\perp}}
  =
  \frac{e^{2}}{4\pi^{3} a}
  \int d^{2} p_{\perp} p_{\perp}^{2}
  K_{0}^{2}(p_{\perp} / a).
\end{equation}
From Eq. 6.576-4 of Ref. \cite{gradshteyn2014table}, we get
\begin{equation}
  \int d^{2}p_{\perp} p_{\perp}^{2} K_{0}^{2}(p_{\perp} / a)
  =
  2\pi \int dp_{\perp} p_{\perp}^{3} K_{0}^{2}(p_{\perp} / a)
  =
  \frac{2\pi a^{4}}{3},
\end{equation}
which yields the final expression
\begin{equation}
  \kappa
  =
  \frac{e^{2} a^{3}}{12\pi^{2}},
\end{equation}
in agreement with the previous results when taking into account the spin degrees of freedom of the underlying field.

\end{appendix}

\bibliographystyle{apsrev4-2}
\bibliography{curvedspace}

\begin{thebibliography}{52}%
\makeatletter
\providecommand \@ifxundefined [1]{%
 \@ifx{#1\undefined}
}%
\providecommand \@ifnum [1]{%
 \ifnum #1\expandafter \@firstoftwo
 \else \expandafter \@secondoftwo
 \fi
}%
\providecommand \@ifx [1]{%
 \ifx #1\expandafter \@firstoftwo
 \else \expandafter \@secondoftwo
 \fi
}%
\providecommand \natexlab [1]{#1}%
\providecommand \enquote  [1]{``#1''}%
\providecommand \bibnamefont  [1]{#1}%
\providecommand \bibfnamefont [1]{#1}%
\providecommand \citenamefont [1]{#1}%
\providecommand \href@noop [0]{\@secondoftwo}%
\providecommand \href [0]{\begingroup \@sanitize@url \@href}%
\providecommand \@href[1]{\@@startlink{#1}\@@href}%
\providecommand \@@href[1]{\endgroup#1\@@endlink}%
\providecommand \@sanitize@url [0]{\catcode `\\12\catcode `\$12\catcode
  `\&12\catcode `\#12\catcode `\^12\catcode `\_12\catcode `\%12\relax}%
\providecommand \@@startlink[1]{}%
\providecommand \@@endlink[0]{}%
\providecommand \url  [0]{\begingroup\@sanitize@url \@url }%
\providecommand \@url [1]{\endgroup\@href {#1}{\urlprefix }}%
\providecommand \urlprefix  [0]{URL }%
\providecommand \Eprint [0]{\href }%
\providecommand \doibase [0]{https://doi.org/}%
\providecommand \selectlanguage [0]{\@gobble}%
\providecommand \bibinfo  [0]{\@secondoftwo}%
\providecommand \bibfield  [0]{\@secondoftwo}%
\providecommand \translation [1]{[#1]}%
\providecommand \BibitemOpen [0]{}%
\providecommand \bibitemStop [0]{}%
\providecommand \bibitemNoStop [0]{.\EOS\space}%
\providecommand \EOS [0]{\spacefactor3000\relax}%
\providecommand \BibitemShut  [1]{\csname bibitem#1\endcsname}%
\let\auto@bib@innerbib\@empty
\bibitem [{\citenamefont {Hawking}(1974)}]{Hawking:1974rv}%
  \BibitemOpen
  \bibfield  {author} {\bibinfo {author} {\bibfnamefont {S.~W.}\ \bibnamefont
  {Hawking}},\ }\href {https://doi.org/10.1038/248030a0} {\bibfield  {journal}
  {\bibinfo  {journal} {Nature}\ }\textbf {\bibinfo {volume} {248}},\ \bibinfo
  {pages} {30} (\bibinfo {year} {1974})}\BibitemShut {NoStop}%
\bibitem [{\citenamefont {Davies}(1975)}]{Davies:1974th}%
  \BibitemOpen
  \bibfield  {author} {\bibinfo {author} {\bibfnamefont {P.~C.~W.}\
  \bibnamefont {Davies}},\ }\href {https://doi.org/10.1088/0305-4470/8/4/022}
  {\bibfield  {journal} {\bibinfo  {journal} {J. Phys. A}\ }\textbf {\bibinfo
  {volume} {8}},\ \bibinfo {pages} {609} (\bibinfo {year} {1975})}\BibitemShut
  {NoStop}%
\bibitem [{\citenamefont {Unruh}(1976)}]{Unruh:1976db}%
  \BibitemOpen
  \bibfield  {author} {\bibinfo {author} {\bibfnamefont {W.~G.}\ \bibnamefont
  {Unruh}},\ }\href {https://doi.org/10.1103/PhysRevD.14.870} {\bibfield
  {journal} {\bibinfo  {journal} {Phys. Rev. D}\ }\textbf {\bibinfo {volume}
  {14}},\ \bibinfo {pages} {870} (\bibinfo {year} {1976})}\BibitemShut
  {NoStop}%
\bibitem [{\citenamefont {Troost}\ and\ \citenamefont
  {Van~Dam}(1977)}]{Troost:1977dw}%
  \BibitemOpen
  \bibfield  {author} {\bibinfo {author} {\bibfnamefont {W.}~\bibnamefont
  {Troost}}\ and\ \bibinfo {author} {\bibfnamefont {H.}~\bibnamefont
  {Van~Dam}},\ }\href {https://doi.org/10.1016/0370-2693(77)90764-X} {\bibfield
   {journal} {\bibinfo  {journal} {Phys. Lett. B}\ }\textbf {\bibinfo {volume}
  {71}},\ \bibinfo {pages} {149} (\bibinfo {year} {1977})}\BibitemShut
  {NoStop}%
\bibitem [{\citenamefont {Troost}\ and\ \citenamefont {van
  Dam}(1979)}]{Troost:1978yk}%
  \BibitemOpen
  \bibfield  {author} {\bibinfo {author} {\bibfnamefont {W.}~\bibnamefont
  {Troost}}\ and\ \bibinfo {author} {\bibfnamefont {H.}~\bibnamefont {van
  Dam}},\ }\href {https://doi.org/10.1016/0550-3213(79)90091-9} {\bibfield
  {journal} {\bibinfo  {journal} {Nucl. Phys. B}\ }\textbf {\bibinfo {volume}
  {152}},\ \bibinfo {pages} {442} (\bibinfo {year} {1979})}\BibitemShut
  {NoStop}%
\bibitem [{\citenamefont {Crispino}\ \emph {et~al.}(2008)\citenamefont
  {Crispino}, \citenamefont {Higuchi},\ and\ \citenamefont
  {Matsas}}]{Crispino:2007eb}%
  \BibitemOpen
  \bibfield  {author} {\bibinfo {author} {\bibfnamefont {L.~C.~B.}\
  \bibnamefont {Crispino}}, \bibinfo {author} {\bibfnamefont {A.}~\bibnamefont
  {Higuchi}},\ and\ \bibinfo {author} {\bibfnamefont {G.~E.~A.}\ \bibnamefont
  {Matsas}},\ }\href {https://doi.org/10.1103/RevModPhys.80.787} {\bibfield
  {journal} {\bibinfo  {journal} {Rev. Mod. Phys.}\ }\textbf {\bibinfo {volume}
  {80}},\ \bibinfo {pages} {787} (\bibinfo {year} {2008})},\ \Eprint
  {https://arxiv.org/abs/0710.5373} {arXiv:0710.5373 [gr-qc]} \BibitemShut
  {NoStop}%
\bibitem [{\citenamefont {Gibbons}\ and\ \citenamefont
  {Perry}(1976)}]{Gibbons:1976es}%
  \BibitemOpen
  \bibfield  {author} {\bibinfo {author} {\bibfnamefont {G.~W.}\ \bibnamefont
  {Gibbons}}\ and\ \bibinfo {author} {\bibfnamefont {M.~J.}\ \bibnamefont
  {Perry}},\ }\href {https://doi.org/10.1103/PhysRevLett.36.985} {\bibfield
  {journal} {\bibinfo  {journal} {Phys. Rev. Lett.}\ }\textbf {\bibinfo
  {volume} {36}},\ \bibinfo {pages} {985} (\bibinfo {year} {1976})},\ \bibinfo
  {note} {; Proc. Roy. Soc. Lond. A, {\bf 358}, 467 (1978)}\BibitemShut
  {NoStop}%
\bibitem [{\citenamefont {Candelas}\ and\ \citenamefont
  {Sciama}(1977)}]{Candelas:1977zz}%
  \BibitemOpen
  \bibfield  {author} {\bibinfo {author} {\bibfnamefont {P.}~\bibnamefont
  {Candelas}}\ and\ \bibinfo {author} {\bibfnamefont {D.~W.}\ \bibnamefont
  {Sciama}},\ }\href {https://doi.org/10.1103/PhysRevLett.38.1372} {\bibfield
  {journal} {\bibinfo  {journal} {Phys. Rev. Lett.}\ }\textbf {\bibinfo
  {volume} {38}},\ \bibinfo {pages} {1372} (\bibinfo {year}
  {1977})}\BibitemShut {NoStop}%
\bibitem [{\citenamefont {Barshay}\ and\ \citenamefont
  {Troost}(1978)}]{Barshay:1977hc}%
  \BibitemOpen
  \bibfield  {author} {\bibinfo {author} {\bibfnamefont {S.}~\bibnamefont
  {Barshay}}\ and\ \bibinfo {author} {\bibfnamefont {W.}~\bibnamefont
  {Troost}},\ }\href {https://doi.org/10.1016/0370-2693(78)90759-1} {\bibfield
  {journal} {\bibinfo  {journal} {Phys. Lett. B}\ }\textbf {\bibinfo {volume}
  {73}},\ \bibinfo {pages} {437} (\bibinfo {year} {1978})}\BibitemShut
  {NoStop}%
\bibitem [{\citenamefont {Kharzeev}\ and\ \citenamefont
  {Tuchin}(2005)}]{Kharzeev:2005iz}%
  \BibitemOpen
  \bibfield  {author} {\bibinfo {author} {\bibfnamefont {D.}~\bibnamefont
  {Kharzeev}}\ and\ \bibinfo {author} {\bibfnamefont {K.}~\bibnamefont
  {Tuchin}},\ }\href {https://doi.org/10.1016/j.nuclphysa.2005.03.001}
  {\bibfield  {journal} {\bibinfo  {journal} {Nucl. Phys. A}\ }\textbf
  {\bibinfo {volume} {753}},\ \bibinfo {pages} {316} (\bibinfo {year}
  {2005})},\ \Eprint {https://arxiv.org/abs/hep-ph/0501234}
  {arXiv:hep-ph/0501234} \BibitemShut {NoStop}%
\bibitem [{\citenamefont {Castorina}\ \emph {et~al.}(2007)\citenamefont
  {Castorina}, \citenamefont {Kharzeev},\ and\ \citenamefont
  {Satz}}]{Castorina:2007eb}%
  \BibitemOpen
  \bibfield  {author} {\bibinfo {author} {\bibfnamefont {P.}~\bibnamefont
  {Castorina}}, \bibinfo {author} {\bibfnamefont {D.}~\bibnamefont
  {Kharzeev}},\ and\ \bibinfo {author} {\bibfnamefont {H.}~\bibnamefont
  {Satz}},\ }\href {https://doi.org/10.1140/epjc/s10052-007-0368-6} {\bibfield
  {journal} {\bibinfo  {journal} {Eur. Phys. J. C}\ }\textbf {\bibinfo {volume}
  {52}},\ \bibinfo {pages} {187} (\bibinfo {year} {2007})},\ \Eprint
  {https://arxiv.org/abs/0704.1426} {arXiv:0704.1426 [hep-ph]} \BibitemShut
  {NoStop}%
\bibitem [{\citenamefont {Biro}\ \emph {et~al.}(2012)\citenamefont {Biro},
  \citenamefont {Gyulassy},\ and\ \citenamefont {Schram}}]{Biro:2011ne}%
  \BibitemOpen
  \bibfield  {author} {\bibinfo {author} {\bibfnamefont {T.~S.}\ \bibnamefont
  {Biro}}, \bibinfo {author} {\bibfnamefont {M.}~\bibnamefont {Gyulassy}},\
  and\ \bibinfo {author} {\bibfnamefont {Z.}~\bibnamefont {Schram}},\ }\href
  {https://doi.org/{10.1016/j.physletb.2011.12.062}} {\bibfield  {journal}
  {\bibinfo  {journal} {Phys. Lett. B}\ }\textbf {\bibinfo {volume} {708}},\
  \bibinfo {pages} {276} (\bibinfo {year} {2012})},\ \Eprint
  {https://arxiv.org/abs/1111.4817} {arXiv:1111.4817 [hep-ph]} \BibitemShut
  {NoStop}%
\bibitem [{\citenamefont {Higuchi}\ \emph
  {et~al.}(1992{\natexlab{a}})\citenamefont {Higuchi}, \citenamefont {Matsas},\
  and\ \citenamefont {Sudarsky}}]{Higuchi:1992we}%
  \BibitemOpen
  \bibfield  {author} {\bibinfo {author} {\bibfnamefont {A.}~\bibnamefont
  {Higuchi}}, \bibinfo {author} {\bibfnamefont {G.~E.~A.}\ \bibnamefont
  {Matsas}},\ and\ \bibinfo {author} {\bibfnamefont {D.}~\bibnamefont
  {Sudarsky}},\ }\href {https://doi.org/10.1103/PhysRevD.45.R3308} {\bibfield
  {journal} {\bibinfo  {journal} {Phys. Rev. D}\ }\textbf {\bibinfo {volume}
  {45}},\ \bibinfo {pages} {R3308} (\bibinfo {year}
  {1992}{\natexlab{a}})}\BibitemShut {NoStop}%
\bibitem [{\citenamefont {Higuchi}\ \emph
  {et~al.}(1992{\natexlab{b}})\citenamefont {Higuchi}, \citenamefont {Matsas},\
  and\ \citenamefont {Sudarsky}}]{Higuchi:1992both}%
  \BibitemOpen
  \bibfield  {author} {\bibinfo {author} {\bibfnamefont {A.}~\bibnamefont
  {Higuchi}}, \bibinfo {author} {\bibfnamefont {G.~E.~A.}\ \bibnamefont
  {Matsas}},\ and\ \bibinfo {author} {\bibfnamefont {D.}~\bibnamefont
  {Sudarsky}},\ }\href {https://doi.org/10.1103/PhysRevD.45.R3308} {\bibfield
  {journal} {\bibinfo  {journal} {Phys. Rev. D}\ }\textbf {\bibinfo {volume}
  {45}},\ \bibinfo {pages} {R3308} (\bibinfo {year} {1992}{\natexlab{b}})},\
  \bibinfo {note} {ibid. 46, 3450 (1992).}\BibitemShut {Stop}%
\bibitem [{\citenamefont {Chen}\ and\ \citenamefont
  {Tajima}(1999)}]{Chen:1998kp}%
  \BibitemOpen
  \bibfield  {author} {\bibinfo {author} {\bibfnamefont {P.}~\bibnamefont
  {Chen}}\ and\ \bibinfo {author} {\bibfnamefont {T.}~\bibnamefont {Tajima}},\
  }\href {https://doi.org/10.1103/PhysRevLett.83.256} {\bibfield  {journal}
  {\bibinfo  {journal} {Phys. Rev. Lett.}\ }\textbf {\bibinfo {volume} {83}},\
  \bibinfo {pages} {256} (\bibinfo {year} {1999})}\BibitemShut {NoStop}%
\bibitem [{\citenamefont {Schutzhold}\ \emph {et~al.}(2006)\citenamefont
  {Schutzhold}, \citenamefont {Schaller},\ and\ \citenamefont
  {Habs}}]{Schutzhold:2006gj}%
  \BibitemOpen
  \bibfield  {author} {\bibinfo {author} {\bibfnamefont {R.}~\bibnamefont
  {Schutzhold}}, \bibinfo {author} {\bibfnamefont {G.}~\bibnamefont
  {Schaller}},\ and\ \bibinfo {author} {\bibfnamefont {D.}~\bibnamefont
  {Habs}},\ }\href {https://doi.org/10.1103/PhysRevLett.97.121302} {\bibfield
  {journal} {\bibinfo  {journal} {Phys. Rev. Lett.}\ }\textbf {\bibinfo
  {volume} {97}},\ \bibinfo {pages} {121302} (\bibinfo {year} {2006})},\
  \bibinfo {note} {[Erratum: Phys.Rev.Lett. 97, 139902 (2006)]},\ \Eprint
  {https://arxiv.org/abs/quant-ph/0604065} {arXiv:quant-ph/0604065}
  \BibitemShut {NoStop}%
\bibitem [{\citenamefont {Schutzhold}\ \emph {et~al.}(2008)\citenamefont
  {Schutzhold}, \citenamefont {Schaller},\ and\ \citenamefont
  {Habs}}]{Schutzhold:2008zza}%
  \BibitemOpen
  \bibfield  {author} {\bibinfo {author} {\bibfnamefont {R.}~\bibnamefont
  {Schutzhold}}, \bibinfo {author} {\bibfnamefont {G.}~\bibnamefont
  {Schaller}},\ and\ \bibinfo {author} {\bibfnamefont {D.}~\bibnamefont
  {Habs}},\ }\href {https://doi.org/10.1103/PhysRevLett.100.091301} {\bibfield
  {journal} {\bibinfo  {journal} {Phys. Rev. Lett.}\ }\textbf {\bibinfo
  {volume} {100}},\ \bibinfo {pages} {091301} (\bibinfo {year} {2008})},\
  \Eprint {https://arxiv.org/abs/0705.4385} {arXiv:0705.4385 [quant-ph]}
  \BibitemShut {NoStop}%
\bibitem [{\citenamefont {Thirolf}\ \emph {et~al.}(2010)\citenamefont {Thirolf}
  \emph {et~al.}}]{Thirolf:2010ozy}%
  \BibitemOpen
  \bibfield  {author} {\bibinfo {author} {\bibfnamefont {P.~G.}\ \bibnamefont
  {Thirolf}} \emph {et~al.},\ }\href {https://doi.org/10.1063/1.3426087}
  {\bibfield  {journal} {\bibinfo  {journal} {AIP Conf. Proc.}\ }\textbf
  {\bibinfo {volume} {1228}},\ \bibinfo {pages} {54} (\bibinfo {year}
  {2010})}\BibitemShut {NoStop}%
\bibitem [{\citenamefont {Anglin}(1993)}]{Anglin:1992ur}%
  \BibitemOpen
  \bibfield  {author} {\bibinfo {author} {\bibfnamefont {J.~R.}\ \bibnamefont
  {Anglin}},\ }\href {https://doi.org/10.1103/PhysRevD.47.4525} {\bibfield
  {journal} {\bibinfo  {journal} {Phys. Rev. D}\ }\textbf {\bibinfo {volume}
  {47}},\ \bibinfo {pages} {4525} (\bibinfo {year} {1993})},\ \Eprint
  {https://arxiv.org/abs/hep-th/9210035} {arXiv:hep-th/9210035} \BibitemShut
  {NoStop}%
\bibitem [{\citenamefont {Johnson}\ and\ \citenamefont
  {Hu}(2002)}]{Johnson:2000qd}%
  \BibitemOpen
  \bibfield  {author} {\bibinfo {author} {\bibfnamefont {P.~R.}\ \bibnamefont
  {Johnson}}\ and\ \bibinfo {author} {\bibfnamefont {B.~L.}\ \bibnamefont
  {Hu}},\ }\href {https://doi.org/10.1103/PhysRevD.65.065015} {\bibfield
  {journal} {\bibinfo  {journal} {Phys. Rev. D}\ }\textbf {\bibinfo {volume}
  {65}},\ \bibinfo {pages} {065015} (\bibinfo {year} {2002})},\ \Eprint
  {https://arxiv.org/abs/quant-ph/0101001} {arXiv:quant-ph/0101001}
  \BibitemShut {NoStop}%
\bibitem [{\citenamefont {Johnson}\ and\ \citenamefont
  {Hu}(2005)}]{Johnson:2005pf}%
  \BibitemOpen
  \bibfield  {author} {\bibinfo {author} {\bibfnamefont {P.~R.}\ \bibnamefont
  {Johnson}}\ and\ \bibinfo {author} {\bibfnamefont {B.~L.}\ \bibnamefont
  {Hu}},\ }\href {https://doi.org/10.1007/s10701-005-6404-1} {\bibfield
  {journal} {\bibinfo  {journal} {Found. Phys.}\ }\textbf {\bibinfo {volume}
  {35}},\ \bibinfo {pages} {1117} (\bibinfo {year} {2005})},\ \Eprint
  {https://arxiv.org/abs/gr-qc/0501029} {arXiv:gr-qc/0501029} \BibitemShut
  {NoStop}%
\bibitem [{\citenamefont {Galley}\ and\ \citenamefont
  {Hu}(2005)}]{Galley:2005tj}%
  \BibitemOpen
  \bibfield  {author} {\bibinfo {author} {\bibfnamefont {C.~R.}\ \bibnamefont
  {Galley}}\ and\ \bibinfo {author} {\bibfnamefont {B.~L.}\ \bibnamefont
  {Hu}},\ }\href {https://doi.org/10.1103/PhysRevD.72.084023} {\bibfield
  {journal} {\bibinfo  {journal} {Phys. Rev. D}\ }\textbf {\bibinfo {volume}
  {72}},\ \bibinfo {pages} {084023} (\bibinfo {year} {2005})},\ \Eprint
  {https://arxiv.org/abs/gr-qc/0505085} {arXiv:gr-qc/0505085} \BibitemShut
  {NoStop}%
\bibitem [{\citenamefont {Galley}\ \emph {et~al.}(2006)\citenamefont {Galley},
  \citenamefont {Hu},\ and\ \citenamefont {Lin}}]{Galley:2006gs}%
  \BibitemOpen
  \bibfield  {author} {\bibinfo {author} {\bibfnamefont {C.~R.}\ \bibnamefont
  {Galley}}, \bibinfo {author} {\bibfnamefont {B.~L.}\ \bibnamefont {Hu}},\
  and\ \bibinfo {author} {\bibfnamefont {S.-Y.}\ \bibnamefont {Lin}},\ }\href
  {https://doi.org/10.1103/PhysRevD.74.024017} {\bibfield  {journal} {\bibinfo
  {journal} {Phys. Rev. D}\ }\textbf {\bibinfo {volume} {74}},\ \bibinfo
  {pages} {024017} (\bibinfo {year} {2006})},\ \Eprint
  {https://arxiv.org/abs/gr-qc/0603099} {arXiv:gr-qc/0603099} \BibitemShut
  {NoStop}%
\bibitem [{\citenamefont {Lin}\ and\ \citenamefont {Hu}(2006)}]{Lin:2005uk}%
  \BibitemOpen
  \bibfield  {author} {\bibinfo {author} {\bibfnamefont {S.-Y.}\ \bibnamefont
  {Lin}}\ and\ \bibinfo {author} {\bibfnamefont {B.}~\bibnamefont {Hu}},\
  }\href {https://doi.org/10.1103/PhysRevD.73.124018} {\bibfield  {journal}
  {\bibinfo  {journal} {Phys. Rev. D}\ }\textbf {\bibinfo {volume} {73}},\
  \bibinfo {pages} {124018} (\bibinfo {year} {2006})},\ \Eprint
  {https://arxiv.org/abs/gr-qc/0507054} {arXiv:gr-qc/0507054} \BibitemShut
  {NoStop}%
\bibitem [{\citenamefont {Lin}\ and\ \citenamefont
  {Hu}(2007{\natexlab{a}})}]{Lin:2006jw}%
  \BibitemOpen
  \bibfield  {author} {\bibinfo {author} {\bibfnamefont {S.-Y.}\ \bibnamefont
  {Lin}}\ and\ \bibinfo {author} {\bibfnamefont {B.}~\bibnamefont {Hu}},\
  }\href {https://doi.org/10.1103/PhysRevD.76.064008} {\bibfield  {journal}
  {\bibinfo  {journal} {Phys. Rev. D}\ }\textbf {\bibinfo {volume} {76}},\
  \bibinfo {pages} {064008} (\bibinfo {year} {2007}{\natexlab{a}})},\ \Eprint
  {https://arxiv.org/abs/gr-qc/0611062} {arXiv:gr-qc/0611062} \BibitemShut
  {NoStop}%
\bibitem [{\citenamefont {Lin}\ and\ \citenamefont
  {Hu}(2007{\natexlab{b}})}]{Lin:2006vz}%
  \BibitemOpen
  \bibfield  {author} {\bibinfo {author} {\bibfnamefont {S.-Y.}\ \bibnamefont
  {Lin}}\ and\ \bibinfo {author} {\bibfnamefont {B.}~\bibnamefont {Hu}},\
  }\href {https://doi.org/10.1007/s10701-007-9120-1} {\bibfield  {journal}
  {\bibinfo  {journal} {Found. Phys.}\ }\textbf {\bibinfo {volume} {37}},\
  \bibinfo {pages} {480} (\bibinfo {year} {2007}{\natexlab{b}})},\ \Eprint
  {https://arxiv.org/abs/gr-qc/0610024} {arXiv:gr-qc/0610024} \BibitemShut
  {NoStop}%
\bibitem [{\citenamefont {Doukas}\ \emph {et~al.}(2013)\citenamefont {Doukas},
  \citenamefont {Lin}, \citenamefont {Hu},\ and\ \citenamefont
  {Mann}}]{Doukas:2013noa}%
  \BibitemOpen
  \bibfield  {author} {\bibinfo {author} {\bibfnamefont {J.}~\bibnamefont
  {Doukas}}, \bibinfo {author} {\bibfnamefont {S.-Y.}\ \bibnamefont {Lin}},
  \bibinfo {author} {\bibfnamefont {B.}~\bibnamefont {Hu}},\ and\ \bibinfo
  {author} {\bibfnamefont {R.~B.}\ \bibnamefont {Mann}},\ }\href
  {https://doi.org/10.1007/JHEP11(2013)119} {\bibfield  {journal} {\bibinfo
  {journal} {JHEP}\ }\textbf {\bibinfo {volume} {11}},\ \bibinfo {pages}
  {119}},\ \Eprint {https://arxiv.org/abs/1307.4360} {arXiv:1307.4360 [gr-qc]}
  \BibitemShut {NoStop}%
\bibitem [{\citenamefont {Iso}\ \emph {et~al.}(2011)\citenamefont {Iso},
  \citenamefont {Yamamoto},\ and\ \citenamefont {Zhang}}]{Iso:2010yq}%
  \BibitemOpen
  \bibfield  {author} {\bibinfo {author} {\bibfnamefont {S.}~\bibnamefont
  {Iso}}, \bibinfo {author} {\bibfnamefont {Y.}~\bibnamefont {Yamamoto}},\ and\
  \bibinfo {author} {\bibfnamefont {S.}~\bibnamefont {Zhang}},\ }\href
  {https://doi.org/10.1103/PhysRevD.84.025005} {\bibfield  {journal} {\bibinfo
  {journal} {Phys. Rev. D}\ }\textbf {\bibinfo {volume} {84}},\ \bibinfo
  {pages} {025005} (\bibinfo {year} {2011})},\ \Eprint
  {https://arxiv.org/abs/1011.4191} {arXiv:1011.4191 [hep-th]} \BibitemShut
  {NoStop}%
\bibitem [{\citenamefont {Bekenstein}(1974)}]{Bekenstein:1974ax}%
  \BibitemOpen
  \bibfield  {author} {\bibinfo {author} {\bibfnamefont {J.~D.}\ \bibnamefont
  {Bekenstein}},\ }\href {https://doi.org/10.1103/PhysRevD.9.3292} {\bibfield
  {journal} {\bibinfo  {journal} {Phys. Rev. D}\ }\textbf {\bibinfo {volume}
  {9}},\ \bibinfo {pages} {3292} (\bibinfo {year} {1974})}\BibitemShut
  {NoStop}%
\bibitem [{\citenamefont {Bekenstein}(1975)}]{Bekenstein:1975tw}%
  \BibitemOpen
  \bibfield  {author} {\bibinfo {author} {\bibfnamefont {J.~D.}\ \bibnamefont
  {Bekenstein}},\ }\href {https://doi.org/10.1103/PhysRevD.12.3077} {\bibfield
  {journal} {\bibinfo  {journal} {Phys. Rev. D}\ }\textbf {\bibinfo {volume}
  {12}},\ \bibinfo {pages} {3077} (\bibinfo {year} {1975})}\BibitemShut
  {NoStop}%
\bibitem [{\citenamefont {Israel}(1976)}]{Israel:1976ur}%
  \BibitemOpen
  \bibfield  {author} {\bibinfo {author} {\bibfnamefont {W.}~\bibnamefont
  {Israel}},\ }\href {https://doi.org/10.1016/0375-9601(76)90178-X} {\bibfield
  {journal} {\bibinfo  {journal} {Phys. Lett. A}\ }\textbf {\bibinfo {volume}
  {57}},\ \bibinfo {pages} {107} (\bibinfo {year} {1976})}\BibitemShut
  {NoStop}%
\bibitem [{\citenamefont {Sciama}\ \emph {et~al.}(1981)\citenamefont {Sciama},
  \citenamefont {Candelas},\ and\ \citenamefont {Deutsch}}]{Sciama:1981hr}%
  \BibitemOpen
  \bibfield  {author} {\bibinfo {author} {\bibfnamefont {D.~W.}\ \bibnamefont
  {Sciama}}, \bibinfo {author} {\bibfnamefont {P.}~\bibnamefont {Candelas}},\
  and\ \bibinfo {author} {\bibfnamefont {D.}~\bibnamefont {Deutsch}},\ }\href
  {https://doi.org/10.1080/00018738100101457} {\bibfield  {journal} {\bibinfo
  {journal} {Adv. Phys.}\ }\textbf {\bibinfo {volume} {30}},\ \bibinfo {pages}
  {327} (\bibinfo {year} {1981})}\BibitemShut {NoStop}%
\bibitem [{\citenamefont {Bell}\ and\ \citenamefont
  {Leinaas}(1983)}]{Bell:1982qr}%
  \BibitemOpen
  \bibfield  {author} {\bibinfo {author} {\bibfnamefont {J.~S.}\ \bibnamefont
  {Bell}}\ and\ \bibinfo {author} {\bibfnamefont {J.~M.}\ \bibnamefont
  {Leinaas}},\ }\href {https://doi.org/10.1016/0550-3213(83)90601-6} {\bibfield
   {journal} {\bibinfo  {journal} {Nucl. Phys. B}\ }\textbf {\bibinfo {volume}
  {212}},\ \bibinfo {pages} {131} (\bibinfo {year} {1983})}\BibitemShut
  {NoStop}%
\bibitem [{\citenamefont {Reif}(1965)}]{reif1965fundamentals}%
  \BibitemOpen
  \bibfield  {author} {\bibinfo {author} {\bibfnamefont {F.}~\bibnamefont
  {Reif}},\ }\href@noop {} {\bibfield  {journal} {\bibinfo  {journal}
  {McGraw-Hill}\ }\textbf {\bibinfo {volume} {5}},\ \bibinfo {pages} {7}
  (\bibinfo {year} {1965})}\BibitemShut {NoStop}%
\bibitem [{\citenamefont {Moore}\ and\ \citenamefont
  {Teaney}(2005)}]{Moore:2004tg}%
  \BibitemOpen
  \bibfield  {author} {\bibinfo {author} {\bibfnamefont {G.~D.}\ \bibnamefont
  {Moore}}\ and\ \bibinfo {author} {\bibfnamefont {D.}~\bibnamefont {Teaney}},\
  }\href {https://doi.org/10.1103/PhysRevC.71.064904} {\bibfield  {journal}
  {\bibinfo  {journal} {Phys. Rev. C}\ }\textbf {\bibinfo {volume} {71}},\
  \bibinfo {pages} {064904} (\bibinfo {year} {2005})},\ \Eprint
  {https://arxiv.org/abs/hep-ph/0412346} {arXiv:hep-ph/0412346} \BibitemShut
  {NoStop}%
\bibitem [{\citenamefont {Coleman}(1982)}]{coleman1982classical}%
  \BibitemOpen
  \bibfield  {author} {\bibinfo {author} {\bibfnamefont {S.}~\bibnamefont
  {Coleman}},\ }in\ \href@noop {} {\emph {\bibinfo {booktitle}
  {Electromagnetism}}}\ (\bibinfo  {publisher} {Springer},\ \bibinfo {year}
  {1982})\ pp.\ \bibinfo {pages} {183--210}\BibitemShut {NoStop}%
\bibitem [{\citenamefont {Nikishov}(1985)}]{nikishov1985problems}%
  \BibitemOpen
  \bibfield  {author} {\bibinfo {author} {\bibfnamefont {A.}~\bibnamefont
  {Nikishov}},\ }\href@noop {} {\bibfield  {journal} {\bibinfo  {journal}
  {Journal of Soviet Laser Research}\ }\textbf {\bibinfo {volume} {6}},\
  \bibinfo {pages} {619} (\bibinfo {year} {1985})}\BibitemShut {NoStop}%
\bibitem [{\citenamefont {Cohen}\ and\ \citenamefont
  {McGady}(2008)}]{Cohen:2008wz}%
  \BibitemOpen
  \bibfield  {author} {\bibinfo {author} {\bibfnamefont {T.~D.}\ \bibnamefont
  {Cohen}}\ and\ \bibinfo {author} {\bibfnamefont {D.~A.}\ \bibnamefont
  {McGady}},\ }\href {https://doi.org/10.1103/PhysRevD.78.036008} {\bibfield
  {journal} {\bibinfo  {journal} {Phys. Rev. D}\ }\textbf {\bibinfo {volume}
  {78}},\ \bibinfo {pages} {036008} (\bibinfo {year} {2008})},\ \Eprint
  {https://arxiv.org/abs/0807.1117} {arXiv:0807.1117 [hep-ph]} \BibitemShut
  {NoStop}%
\bibitem [{\citenamefont {Higuchi}\ and\ \citenamefont
  {Matsas}(1993)}]{Higuchi:1993fn}%
  \BibitemOpen
  \bibfield  {author} {\bibinfo {author} {\bibfnamefont {A.}~\bibnamefont
  {Higuchi}}\ and\ \bibinfo {author} {\bibfnamefont {G.~E.~A.}\ \bibnamefont
  {Matsas}},\ }\href {https://doi.org/10.1103/PhysRevD.48.689} {\bibfield
  {journal} {\bibinfo  {journal} {Phys. Rev. D}\ }\textbf {\bibinfo {volume}
  {48}},\ \bibinfo {pages} {689} (\bibinfo {year} {1993})}\BibitemShut
  {NoStop}%
\bibitem [{\citenamefont {Calzetta}\ and\ \citenamefont
  {Hu}(2008)}]{Calzetta:2008iqa}%
  \BibitemOpen
  \bibfield  {author} {\bibinfo {author} {\bibfnamefont {E.~A.}\ \bibnamefont
  {Calzetta}}\ and\ \bibinfo {author} {\bibfnamefont {B.-L.~B.}\ \bibnamefont
  {Hu}},\ }\href {https://doi.org/10.1017/CBO9780511535123} {\emph {\bibinfo
  {title} {{Nonequilibrium Quantum Field Theory}}}},\ Cambridge Monographs on
  Mathematical Physics\ (\bibinfo  {publisher} {Cambridge University Press},\
  \bibinfo {year} {2008})\BibitemShut {NoStop}%
\bibitem [{Note1()}]{Note1}%
  \BibitemOpen
  \bibinfo {note} {Note however that in \cite {Iso:2010yq}, an incorrect
  definition of $D$ in Eq. 3.23 leads to a result differing by a factor 2 from
  ours.}\BibitemShut {Stop}%
\bibitem [{\citenamefont {Heisenberg}\ and\ \citenamefont
  {Euler}(1936)}]{Heisenberg:1936nmg}%
  \BibitemOpen
  \bibfield  {author} {\bibinfo {author} {\bibfnamefont {W.}~\bibnamefont
  {Heisenberg}}\ and\ \bibinfo {author} {\bibfnamefont {H.}~\bibnamefont
  {Euler}},\ }\href {https://doi.org/10.1007/BF01343663} {\bibfield  {journal}
  {\bibinfo  {journal} {Z. Phys.}\ }\textbf {\bibinfo {volume} {98}},\ \bibinfo
  {pages} {714} (\bibinfo {year} {1936})},\ \Eprint
  {https://arxiv.org/abs/physics/0605038} {arXiv:physics/0605038} \BibitemShut
  {NoStop}%
\bibitem [{\citenamefont {Pauchy~Hwang}\ and\ \citenamefont
  {Kim}(2009)}]{PauchyHwang:2009rz}%
  \BibitemOpen
  \bibfield  {author} {\bibinfo {author} {\bibfnamefont {W.-Y.}\ \bibnamefont
  {Pauchy~Hwang}}\ and\ \bibinfo {author} {\bibfnamefont {S.~P.}\ \bibnamefont
  {Kim}},\ }\href {https://doi.org/10.1103/PhysRevD.80.065004} {\bibfield
  {journal} {\bibinfo  {journal} {Phys. Rev. D}\ }\textbf {\bibinfo {volume}
  {80}},\ \bibinfo {pages} {065004} (\bibinfo {year} {2009})},\ \Eprint
  {https://arxiv.org/abs/0906.3813} {arXiv:0906.3813 [hep-th]} \BibitemShut
  {NoStop}%
\bibitem [{\citenamefont {Labun}\ and\ \citenamefont
  {Rafelski}(2012)}]{Labun:2012jf}%
  \BibitemOpen
  \bibfield  {author} {\bibinfo {author} {\bibfnamefont {L.}~\bibnamefont
  {Labun}}\ and\ \bibinfo {author} {\bibfnamefont {J.}~\bibnamefont
  {Rafelski}},\ }\href {https://doi.org/10.1103/PhysRevD.86.041701} {\bibfield
  {journal} {\bibinfo  {journal} {Phys. Rev. D}\ }\textbf {\bibinfo {volume}
  {86}},\ \bibinfo {pages} {041701} (\bibinfo {year} {2012})},\ \Eprint
  {https://arxiv.org/abs/1203.6148} {arXiv:1203.6148 [hep-ph]} \BibitemShut
  {NoStop}%
\bibitem [{\citenamefont {Svetitsky}(1988)}]{Svetitsky:1987gq}%
  \BibitemOpen
  \bibfield  {author} {\bibinfo {author} {\bibfnamefont {B.}~\bibnamefont
  {Svetitsky}},\ }\href {https://doi.org/10.1103/PhysRevD.37.2484} {\bibfield
  {journal} {\bibinfo  {journal} {Phys. Rev. D}\ }\textbf {\bibinfo {volume}
  {37}},\ \bibinfo {pages} {2484} (\bibinfo {year} {1988})}\BibitemShut
  {NoStop}%
\bibitem [{\citenamefont {Braaten}\ and\ \citenamefont
  {Thoma}(1991)}]{Braaten:1991jj}%
  \BibitemOpen
  \bibfield  {author} {\bibinfo {author} {\bibfnamefont {E.}~\bibnamefont
  {Braaten}}\ and\ \bibinfo {author} {\bibfnamefont {M.~H.}\ \bibnamefont
  {Thoma}},\ }\href {https://doi.org/10.1103/PhysRevD.44.1298} {\bibfield
  {journal} {\bibinfo  {journal} {Phys. Rev. D}\ }\textbf {\bibinfo {volume}
  {44}},\ \bibinfo {pages} {1298} (\bibinfo {year} {1991})},\ \bibinfo {note}
  {ibid. {\bf 44}, R2625 (1991)}\BibitemShut {NoStop}%
\bibitem [{\citenamefont {Labun}\ and\ \citenamefont
  {Rafelski}(2009)}]{Labun:2009vdt}%
  \BibitemOpen
  \bibfield  {author} {\bibinfo {author} {\bibfnamefont {L.}~\bibnamefont
  {Labun}}\ and\ \bibinfo {author} {\bibfnamefont {J.}~\bibnamefont
  {Rafelski}},\ }\href {https://doi.org/10.1103/PhysRevD.79.057901} {\bibfield
  {journal} {\bibinfo  {journal} {Phys. Rev. D}\ }\textbf {\bibinfo {volume}
  {79}},\ \bibinfo {pages} {057901} (\bibinfo {year} {2009})}\BibitemShut
  {NoStop}%
\bibitem [{\citenamefont {Casalderrey-Solana}\ and\ \citenamefont
  {Teaney}(2006)}]{Casalderrey-Solana:2006fio}%
  \BibitemOpen
  \bibfield  {author} {\bibinfo {author} {\bibfnamefont {J.}~\bibnamefont
  {Casalderrey-Solana}}\ and\ \bibinfo {author} {\bibfnamefont
  {D.}~\bibnamefont {Teaney}},\ }\href
  {https://doi.org/10.1103/PhysRevD.74.085012} {\bibfield  {journal} {\bibinfo
  {journal} {Phys. Rev. D}\ }\textbf {\bibinfo {volume} {74}},\ \bibinfo
  {pages} {085012} (\bibinfo {year} {2006})},\ \Eprint
  {https://arxiv.org/abs/hep-ph/0605199} {arXiv:hep-ph/0605199} \BibitemShut
  {NoStop}%
\bibitem [{\citenamefont {Jackson}(1998)}]{Jackson:1998nia}%
  \BibitemOpen
  \bibfield  {author} {\bibinfo {author} {\bibfnamefont {J.~D.}\ \bibnamefont
  {Jackson}},\ }\href@noop {} {\emph {\bibinfo {title} {{Classical
  Electrodynamics}}}}\ (\bibinfo  {publisher} {Wiley},\ \bibinfo {year}
  {1998})\BibitemShut {NoStop}%
\bibitem [{\citenamefont {Gradshteyn}\ and\ \citenamefont
  {Ryzhik}(2014)}]{gradshteyn2014table}%
  \BibitemOpen
  \bibfield  {author} {\bibinfo {author} {\bibfnamefont {I.~S.}\ \bibnamefont
  {Gradshteyn}}\ and\ \bibinfo {author} {\bibfnamefont {I.~M.}\ \bibnamefont
  {Ryzhik}},\ }\href@noop {} {\emph {\bibinfo {title} {Table of integrals,
  series, and products}}}\ (\bibinfo  {publisher} {Academic press},\ \bibinfo
  {year} {2014})\BibitemShut {NoStop}%
\bibitem [{\citenamefont {Anderson}\ and\ \citenamefont
  {Mottola}(2014)}]{Anderson:2013both}%
  \BibitemOpen
  \bibfield  {author} {\bibinfo {author} {\bibfnamefont {P.~R.}\ \bibnamefont
  {Anderson}}\ and\ \bibinfo {author} {\bibfnamefont {E.}~\bibnamefont
  {Mottola}},\ }\href {https://doi.org/10.1103/PhysRevD.89.104038} {\bibfield
  {journal} {\bibinfo  {journal} {Phys. Rev. D}\ }\textbf {\bibinfo {volume}
  {89}},\ \bibinfo {pages} {104038} (\bibinfo {year} {2014})},\ \bibinfo {note}
  {\emph{ibid.} Phys. Rev. D, 89, 104039 (2014)},\ \Eprint
  {https://arxiv.org/abs/1310.0030} {arXiv:1310.0030 [gr-qc]} \BibitemShut
  {NoStop}%
\bibitem [{\citenamefont {Peskin}\ and\ \citenamefont
  {Schroeder}(1995)}]{Peskin:1995ev}%
  \BibitemOpen
  \bibfield  {author} {\bibinfo {author} {\bibfnamefont {M.~E.}\ \bibnamefont
  {Peskin}}\ and\ \bibinfo {author} {\bibfnamefont {D.~V.}\ \bibnamefont
  {Schroeder}},\ }\href@noop {} {\emph {\bibinfo {title} {{An Introduction to
  quantum field theory}}}}\ (\bibinfo  {publisher} {Addison-Wesley},\ \bibinfo
  {address} {Reading, USA},\ \bibinfo {year} {1995})\BibitemShut {NoStop}%
\end{thebibliography}%

\end{document}